\documentclass[12pt]{article}
\usepackage{amssymb, latexsym,amsmath,graphicx,epstopdf,dsfont,outlines,tikz,placeins,amsthm,changepage,relsize,subfig}
\usepackage[margin=1in]{geometry}

\newcommand{\argmax}[1]{\underset{#1}{\operatorname{arg}\,\operatorname{max}}\;}

\makeatletter
\renewcommand\@biblabel[1]{}
\makeatother

\begin{document}
\textwidth 6.5in

\begin{center}
{\Large\bf Model-Based Longitudinal Clustering with Varying Cluster Assignments}

\vspace{0.5in}

Daniel K. Sewell,  Yuguo Chen, William Bernhard and Tracy Sulkin
\footnote{Daniel K. Sewell is Ph.D Candidate, Department of Statistics,
University of Illinois at Urbana-Champaign, Champaign, IL 61820 (E-mail: {\it dsewell2@illinois.edu}). Yuguo Chen
is Associate Professor, Department of Statistics, University of Illinois at
Urbana-Champaign, Champaign, IL 61820 (E-mail: {\it yuguo@illinois.edu}).
Tracy Sulkin is Professor, Department of Political Science, University of Illinois at Urbana-Champaign, Urbana, IL 61801 (E-mail: {\it tsulkin@illinois.edu}).
William Bernhard is Professor, Department of Political Science, University of Illinois at Urbana-Champaign, Urbana, IL 61801 (E-mail: {\it bernhard@illinois.edu}).
This work was supported in part by National Science Foundation grants DMS-11-06796 and DMS-14-06455.}
\end{center}

\begin{abstract}
It is often of interest to perform clustering on longitudinal data, yet it is difficult to formulate an intuitive model for which estimation is computationally feasible.  We propose a model-based clustering method for clustering objects that are observed over time.  The proposed model can be viewed as an extension of the normal mixture model for clustering to longitudinal data.  While existing models only account for clustering effects, we propose modeling the distribution of the observed values of each object as a blending of a cluster effect and an individual effect, hence also giving an estimate of how much the behavior of an object is determined by the cluster to which it belongs.  Further, it is important to detect how explanatory variables affect the clustering.  An advantage of our method is that it can handle multiple explanatory variables of any type through a linear modeling of the cluster transition probabilities.  We implement the generalized EM algorithm using several recursive relationships to greatly decrease the computational cost.  The accuracy of our estimation method is illustrated in a simulation study, and U.S. Congressional data is analyzed.
\vspace{ 2mm}


\end{abstract}

\newpage

\section{Introduction}
Numerous applications exist in which it is of interest to partition the data into homogeneous groups, or clusters.  With longitudinal data in which we observe many objects over a series of time points, one often wishes to better understand how and why these objects are grouped and how they transition from one group to another over time.  Longitudinal clustering is an important topic in such fields as genomics, clinical research, political science, etc.

Much of the previous work on clustering longitudinal data focuses on curve clustering, also called trajectory clustering.  Curve clustering methods view the time series as curves or sometimes as random functions, and intend to give cluster assignments of the objects based on shape similarity.  For example, Ray and Mallick (2006) used Bayesian wavelets to model the curves, and Luan and Li (2003) used B-splines to model the cluster means; often this type of longitudinal clustering is achieved by regression-based approaches (see, e.g., Lou, Jiang, and Keng (1993); Gaffney and Smyth (1999)).

More related to our work is latent transition analysis (LTA) (see, e.g., Graham et al. (1991); Collins and Wugalter (1992); Collins et al. (1994)), 
which is commonly used in the social sciences.  LTA assumes that observed categorical responses are noisy measurements of sequential latent stages; i.e., we can imperfectly measure over time a subject's latent stage (at each time point).  Vermunt, Langeheine, and Bockenholt (1999) and Chung, Park, and Lanza (2005) were able to incorporate extra covariates in their model to help estimate the latent stages.  However, these methods are focused on categorical panel data and sequential stages hypothesized a priori which are strongly associated with the measured categorical variables.

We are concerned with clustering vectors of continuous data observed over time, and where the clusters are based on locations rather than the shape of the trajectories, i.e., we cluster based on the objects' observed values rather than the shape of the objects' time series.  A simple algorithm for this context is the extension of the k-means method for longitudinal data (Genolini and Falissard (2010)).  This does not, however, account for the correlation between each object at different time points, but simply vectorizes each object's time series, and then performs k-means based on the Euclidean distance (or some other metric) of the vectorized time series.  Further, this method requires the objects' time series to all be of the same length, which may not be the case in practice.  A model based approach which accounts for correlation across time points was given by De la Cruz-Mes\'ia, Quintana, and Marshall (2008), who extended the model consisting of mixture of normal distributions (see Banfield and Raftery (1993)) to univariate time series data.  Further work in the context of mixture of normal distributions was done by McNicholas and Murphy (2010), who derived estimates for specific covariance constraints corresponding to univariate time series data.  Anderlucci and Viroli (2014) built on these works to handle multivariate data.  

Scott, James, and Sugar (2005) used a hidden Markov model to estimate the cluster assignments which were allowed to differ for each object at each time point, estimation done using Markov chain Monte Carlo methods.  While a vast improvement, this work still assumes local independence, ignoring individual effects that may be important in modeling the data, and is not flexible enough to handle multiple explanatory variables of varying types.

Here we consider  data consisting of multiple variables (the response variables) at many (and possibly at a varying number of) time points from many objects.  We wish to cluster these objects at each time point based on the similarity of their responses.  The number of clusters is assumed to be fixed, and the clusters themselves are assumed to be static in the sense that the structures that define each cluster are constant over time.  These defining structures will be described in Section \ref{Models}.  We propose a model-based clustering algorithm for longitudinal data that allows cluster assignments to change over time and borrows information across each object's time series when making these assignments, thus allowing the researcher to better understand how these objects transition from one group to another.  We incorporate temporal dependence into the model by modeling each object's current location as a blending of the current (unobserved) cluster assignment and the object's previous values.  
Our model allows explanatory variables of any form to be incorporated into the model in order to explain the clustering by incorporating multinomial logistic regression into the model.  This differs from such other clustering models that incorporate covariates as those of De la Cruz-Mes\'ia et al. (2008) and Anderlucci and Viroli (2014), in that the group memberships of the objects are predicted by the covariates.  This results in clustering being performed purely on the response variables while simultaneously learning how the covariates explain the clustering results.  Estimation is accomplished by using a generalized EM algorithm.  The computational cost of a straightforward implementation of this algorithm can be, however, prohibitively high.  To address this, we derive several recursive relationships which greatly reduce the computational cost from exponential to linear with respect to the number of time points.

The rest of the paper is organized as follows.  Section 2 describes the proposed model.  Section 3 outlines the details of the generalized EM algorithm.  Section 4 provides the results of a simulation study.  Section 5 gives the analysis of longitudinal data collected on Democrats serving in the U.S. House of Representatives.  Section 6 gives a brief discussion.

\section{Models}
\label{Models}
We first present in Section \ref{Nonregressedclusterprobabilities} the basic model for clustering longitudinal data.  We then extend this model in Section \ref{Regressedclusterprobabilities} to model how covariate information may explain the clusters of the response variables.  One can then partition the data into clusters and discover how covariates can explain the clustering assignments.

\subsection{Non-regressed cluster probabilities}
\label{Nonregressedclusterprobabilities}
Denote the value of the $i^{th}$ object at time $t$ as ${\bf X}_{it}$, and let ${\cal X}_i=({\bf X}_{i1}', \ldots, {\bf X}_{iT_i}')'$, for $i=1,\ldots, n$, denote the data to be clustered, where $n$ is the number of distinct objects, $T_i$ is the length of the $i^{th}$ time series, and the dimension of each ${\bf X}_{it}$ is $p$.  We assume that each object is initially assigned to one of $K$ clusters according to the $1\times K$ probability vector $\boldsymbol\alpha=(\alpha_1,\ldots,\alpha_K)$ ($K$ is a fixed positive integer assumed to be known).  The notation to be used throughout lets $Z_{it}$ be the $i^{th}$ object's cluster assignment at time $t$.  Hence $\mathbb{P}(Z_{i1}=k)=\alpha_k$.  For subsequent time points, these objects are assigned to a cluster according to the $K\times K$ transition matrix $\boldsymbol\beta$, $\mathbb{P}(Z_{it}=k|Z_{i(t-1)}=h)=\beta_{hk}$.

Hidden Markov models are in wide use (Fr\"uhwirth-Schnatter (2006)).  Such usefulness motivates us to borrow principles from this class of models.    As will be seen, each cluster in our context is defined by a specific static probability distribution.  A hidden Markov model applied to our setting implies that if we knew which cluster a particular object belonged to at each time point, the observed temporal observations from the object would look like random samples from the clusters' corresponding distributions.  In practice one would expect these observations to still be dependent in some way, even after accounting for the clustering.  For example, in the context of clustering human behaviors, we might expect a person's future behavior to depend on that person's past behavior as well as on the influence exerted on him/her by the group to which he/she belongs.  Then the current expected behavior of a person should be some blending of previous behavior (individual effect) and the overall expected behavior from someone coming from the group to which the person belongs (cluster effect).  This more general type of dependency is known as a Markov switching model.

We make two assumptions about the dependence structure of the model:  given all past cluster assignments, the current cluster assignment depends only on the cluster assignment at the previous time point; given the entire history of both the cluster assignments and the object's values, the current value of the object depends only on the value of the object at the previous time point as well as the current cluster assignment.  These requirements amount to the dependency assumptions
\begin{equation}
Z_{it}|Z_{i1},\ldots,Z_{i(t-1)}\stackrel{{\cal D}}{=}Z_{it}|Z_{i(t-1)}\triangleq \beta_{Z_{i(t-1)}Z_{it}},
\label{Zdepend}
\end{equation}
(hence the cluster assignments follow a Markov process) and
\begin{equation}
{\bf X}_{it}|{\bf X}_{i1},\ldots,{\bf X}_{i(t-1)},Z_{i1},\ldots,Z_{iT_i}\stackrel{{\cal D}}{=} {\bf X}_{it}|{\bf X}_{i(t-1)},Z_{it}.
\label{Xdepend}
\end{equation}

A widely used clustering method is to fit a normal mixture model, and the commonly used k-means is but a special case of this.  This approach performs well in a wide range of settings and so we make the distributional assumptions that
\begin{eqnarray}
\pi({\bf X}_{i1}|Z_{i1}) &=&N({\bf X}_{i1}|\boldsymbol{\mu}_{Z_{i1}},\Sigma_{Z_{i1}}), \label{transition1} \\
\pi({\bf X}_{it}|{\bf X}_{i(t-1)},Z_{it}) &=&N({\bf X}_{it}|\lambda\boldsymbol{\mu}_{Z_{it}}+(1-\lambda){\bf X}_{i(t-1)},\Sigma_{Z_{it}}),
\label{transition2}
\end{eqnarray}
where $\boldsymbol\mu_k$ and $\Sigma_k$ are the mean vector and covariance matrix for the $k^{th}$ cluster for $k=1,\ldots,K$, $\lambda\in (0,1)$, and $N({\bf x}|\boldsymbol\mu,\Sigma)$ denotes the multivariate normal density with mean $\boldsymbol\mu$ and covariance matrix $\Sigma$.  The $\boldsymbol{\mu}_k$'s and $\Sigma_k$'s are the temporally constant structures which define each cluster.  The transition distribution in (\ref{transition2}) blends the role of the current cluster with the individual effect.  That is, in this framework one can look at the distribution of the current position of the $i^{th}$ individual as being influenced by where the individual has been, and by a sense of belonging to a particular cluster.  This model can be thought of as extending the normal mixture model clustering to longitudinal data.

Assuming independence between the objects to be clustered, the complete-data likelihood (the distribution of the observed data and unobserved cluster assignments) can be written as
\begin{equation}
\prod_{i=1}^n\pi({\bf X}_{i1},\ldots,{\bf X}_{iT_i}, Z_{i1},\ldots Z_{iT_i})
=\prod_{i=1}^n\alpha_{Z_{i1}}\pi({\bf X}_{i1}|Z_{i1})\prod_{t=2}^{T_i}\beta_{Z_{i(t-1)}Z_{it}}\pi({\bf X}_{it}|{\bf X}_{i(t-1)},Z_{it}).
\label{completeLik}
\end{equation}
We can then write the marginal distribution of the data as
\footnotesize\begin{align}\nonumber
&\prod_{i=1}^n\pi({\bf X}_{i1},\ldots,{\bf X}_{iT_i}) =&\\
&\prod_{i=1}^n \sum_{Z_{i1}=1}^K\hspace{-0.5pc}\Big( \alpha_{\mathsmaller{Z_{i1}}}
\pi({\bf X}_{i1}|Z_{i1})
\sum_{Z_{i2}=1}^K\hspace{-0.5pc}\Big( \beta_{\mathsmaller{Z_{i1}Z_{i2}}}\pi({\bf X}_{i2}|{\bf X}_{i1},Z_{i2})
\cdots \hspace{-0.5pc}
\sum_{Z_{iT_i}=1}^K \hspace{-0.5pc}\Big(\beta_{\mathsmaller{Z_{i(T_i-1)}Z_{iT_i}}}\pi({\bf X}_{iT_i}|{\bf X}_{i(T_i-1)},Z_{iT_i})\Big)\cdots\hspace{-0.25pc}\Big).&
\label{Xmarginal}
\end{align}\normalsize

\subsection{Regressed cluster probabilities}
\label{Regressedclusterprobabilities}
Suppose there is a set of explanatory variables that may influence how the objects are clustered.  That is, we are not interested in clustering the objects based on the values of these explanatory variables, but rather we are interested in how these variables explain the clustering results.  Let ${\bf w}_{it}$ denote a vector of length $d$ corresponding to these explanatory variables for the $i^{th}$ object at time $t$.  Then the model described in Section \ref{Nonregressedclusterprobabilities} can be slightly modified to allow $\boldsymbol\alpha$ and $\boldsymbol\beta$ to be functions of the explanatory variables, ${\bf w}_{it}$, and unknown parameters, denoted as $\delta_{\ell k}$ and $\boldsymbol\gamma_{\ell k}$ for $\ell=0,\ldots,K$ and $k=1,\ldots,K$.  Mimicking the multinomial logistic regression model, for $k=1,\ldots,K$ we let
\begin{equation}
\log\left(\frac{\alpha_k({\bf w}_{it})}{\alpha_K({\bf w}_{it})} \right)=\delta_{0k}+{\bf w}_{it}'\boldsymbol\gamma_{0k},
\end{equation}
where we fix $\delta_{0K}=0$ and $\boldsymbol\gamma_{0K}={\bf 0}$ for identifiability; similarly for each $h=1,\ldots,K$, for $k=1,\ldots,K$ we let
\begin{equation}
\log\left(\frac{\beta_{hk}({\bf w}_{it})}{\beta_{hK}({\bf w}_{it})} \right)=\delta_{hk}+{\bf w}_{it}'\boldsymbol\gamma_{hk},
\end{equation}
where again we fix $\delta_{hK}=0$ and $\boldsymbol\gamma_{hK}={\bf 0}$ for $h=1,\ldots,K$.  Then we can write the initial and transition cluster probabilities, respectively, as
\begin{eqnarray}
\alpha_{k}({\bf w}_{it})=\frac{\exp(\delta_{0k}+{\bf w}_{it}'\boldsymbol\gamma_{0k})}{1+\sum_{\ell=1}^{K-1}\exp(\delta_{0\ell}+{\bf w}_{it}'\boldsymbol\gamma_{0\ell})} \\
\beta_{hk}({\bf w}_{it})=\frac{\exp(\delta_{hk}+{\bf w}_{it}'\boldsymbol\gamma_{hk})}{1+\sum_{\ell=1}^{K-1}\exp(\delta_{h\ell}+{\bf w}_{it}'\boldsymbol\gamma_{h\ell})}
\end{eqnarray}
for $h,k=1,\ldots,K$.  It is easy to see that setting $\boldsymbol\gamma_{0k}=\boldsymbol\gamma_{hk}={\bf 0}$ for all $h,k$ yields a model equivalent to the non-regressed transition model of Section 2.1.

The model of Section 2.1 then changes in that we are conditioning on the explanatory variables ${\bf w}_{it}$.  Specifically we have that the conditional distribution of the cluster assignments changes from (\ref{Zdepend}) to
\begin{equation}
Z_{it}|Z_{i1},\ldots,Z_{i(t-1)},{\bf w}_{i1},\ldots,{\bf w}_{iT_i}\stackrel{{\cal D}}{=}Z_{it}|Z_{i(t-1)},{\bf w}_{it}\triangleq \beta_{Z_{i(t-1)}Z_{it}}({\bf w}_{it}),
\end{equation}
while (\ref{Xdepend}) does not change.  The complete-data likelihood is
\begin{align}\nonumber
&\prod_{i=1}^n\pi({\bf X}_{i1},\ldots,{\bf X}_{iT_i}, Z_{i1},\ldots Z_{iT_i}|{\bf w}_{i1},\ldots,{\bf w}_{iT_i})& \\
=&\prod_{i=1}^n\alpha_{\mathsmaller{Z_{i1}}}({\bf w}_{i1})\pi({\bf X}_{i1}|Z_{i1})\prod_{t=2}^{T_i}\beta_{\mathsmaller{Z_{i(t-1)}Z_{it}}}({\bf w}_{it})\pi({\bf X}_{it}|{\bf X}_{i(t-1)},Z_{it}),&\vspace{-1pc}
\label{completeLikReg}
\end{align}
and the marginal likelihood is
\begin{align}\nonumber
&\prod_{i=1}^n\pi({\bf X}_{i1},\ldots,{\bf X}_{iT_i}|{\bf w}_{i1},\ldots,{\bf w}_{iT_i}) &\\ \nonumber
=&\prod_{i=1}^n \sum_{Z_{i1}=1}^K\hspace{-0.5pc}\Big( \alpha_{\mathsmaller{Z_{i1}}}({\bf w}_{i1})
\pi({\bf X}_{i1}|Z_{i1})
\sum_{Z_{i2}=1}^K\hspace{-0.5pc}\Big( \beta_{\mathsmaller{Z_{i1}Z_{i2}}}({\bf w}_{i2})\pi({\bf X}_{i2}|{\bf X}_{i1},Z_{i2}) & \\
& \cdots  \sum_{Z_{iT_i}=1}^K \hspace{-0.5pc}\Big(\beta_{\mathsmaller{Z_{i(T_i-1)}Z_{iT_i}}}({\bf w}_{iT_i})\pi({\bf X}_{iT_i}|{\bf X}_{i(T_i-1)},Z_{iT_i})\Big)\cdots\hspace{-0.25pc}\Big).&
\end{align}

\subsection{Computational issues}
The estimation procedure to be outlined in Section \ref{Estimation} is an iterative procedure to maximize the likelihood, the marginal distribution of the data (\ref{Xmarginal}), and hence we must be able to compute the likelihood to know if we have reached convergence.  In computing the likelihood we run into a problem --- the number of terms to be summed grows exponentially with $T_i$.  To make this less computationally expensive, we can consider two recursive relationships.  First consider the joint density $\pi({\bf X}_1,\ldots,{\bf X}_t,Z_t)$ (where the subscript $i$ is suppressed for ease of notation where obvious).  It can be shown that $\pi(Z_t|Z_{t-1},{\bf X}_1,\ldots,{\bf X}_{t-1})=\beta_{Z_{t-1}Z_t}$ (see Appendix \ref{betaZtZtm1}).  Using this, we see that
\begin{align}\nonumber
\pi({\bf X}_1,\ldots,{\bf X}_t,Z_t) &=\hspace{-0.5pc} \sum_{Z_{t-1}=1}^K \pi({\bf X}_t|{\bf X}_{t-1},Z_t)\pi(Z_{t-1},Z_t|{\bf X}_1,\ldots,{\bf X}_{t-1})\pi({\bf X}_1,\ldots,{\bf X}_{t-1})& \\
&=\pi({\bf X}_1,\ldots,{\bf X}_{t-1})\hspace{-0.5pc} \sum_{Z_{t-1}=1}^K \pi(Z_{t-1}|{\bf X}_1,\ldots,{\bf X}_{t-1}) \beta_{\mathsmaller{Z_{t-1}Z_t}} \pi({\bf X}_t|{\bf X}_{t-1},Z_t).
\label{x1totZt}
\end{align}
Computing the marginal likelihood $\pi({\bf X}_1,\ldots,{\bf X}_T)$ can be accomplished by using this recursion in two steps, first obtaining  $\pi(Z_{t-1}|{\bf X}_1,\ldots,{\bf X}_{t-1})$ and then $\pi({\bf X}_1,\ldots,{\bf X}_t,Z_t)$.  More specifically, we first compute and normalize
\begin{equation}
\pi(Z_1|{\bf X}_1) \propto \pi({\bf X}_1|Z_1)\alpha_{Z_1}.
\label{z1givenx1}
\end{equation}  Then for $t=3,\ldots,T$, from (\ref{x1totZt}) we can compute and normalize
\begin{eqnarray}
\pi(Z_{t-1}|{\bf X}_1,\ldots,{\bf X}_{t-1}) \propto&\sum_{Z_{t-2}=1}^K \pi(Z_{t-2}|{\bf X}_1,\ldots,{\bf X}_{t-2}) \beta_{\mathsmaller{Z_{t-2}Z_{t-1}}} \pi({\bf X}_{t-1}|{\bf X}_{t-2},Z_{t-1}).
\label{ztm1givenx}
\end{eqnarray}
Second, we can compute $\pi({\bf X}_1)=\sum_{Z_1=1}^K\pi({\bf X}_1|Z_1)\alpha_{Z_1}$, and then use the normalized results from (\ref{z1givenx1}) and (\ref{ztm1givenx}) to compute, for $t=2,\ldots,T$,
\begin{eqnarray}
\pi({\bf X}_1,\ldots,{\bf X}_t)
=\pi({\bf X}_1,\ldots,{\bf X}_{t-1})\hspace{-0.5pc} \sum_{Z_t=1}^K\sum_{Z_{t-1}=1}^K \pi(Z_{t-1}|{\bf X}_1,\ldots,{\bf X}_{t-1}) \beta_{\mathsmaller{Z_{t-1}Z_t}} \pi({\bf X}_t|{\bf X}_{t-1},Z_t).
\label{endCompRef}
\end{eqnarray}
By utilizing these recursive relationships, the number of terms required to be summed in the computation of the likelihood density grows linearly, rather than exponentially, in time.  Regarding the context of regressed cluster assignment probabilities, the previous work can still be applied simply by exchanging $\alpha_{Z_{i1}}$ for $\alpha_{Z_{i1}({\bf w}_{i1})}$ and $\beta_{Z_{i(t-1)}Z_{it}}$ for $\beta_{Z_{i(t-1)}Z_{it}}({\bf w}_{it})$ in equations (\ref{x1totZt}) to (\ref{endCompRef}).

\section{Estimation}
\label{Estimation}
We aim to find the maximum likelihood estimators (MLE's) for the model parameters, henceforth denoted as $\Theta$.  In the case of non-regressed cluster probabilities, $$\Theta=\{ \boldsymbol\mu_1,\ldots,\boldsymbol\mu_K,\Sigma_1,\ldots,\Sigma_K,\lambda,\alpha_1,\ldots,\alpha_K,\beta_{11},\ldots,\beta_{KK} \},$$ and for the case of regressed cluster probabilities, $$\Theta=\{ \boldsymbol\mu_1,\ldots,\boldsymbol\mu_K,\Sigma_1,\ldots,\Sigma_K,\lambda, \delta_{01},\ldots,\delta_{K(K-1)},\boldsymbol\gamma_{01},\ldots,\boldsymbol\gamma_{K(K-1)} \}.$$

To find the MLE's, we employ the generalized EM algorithm (Dempster, Laird, and Rubin (1977); Wu (1983)) to obtain parameter estimates, as well as cluster assignment probabilities.  Heretofore the notation for probability densities have not included the parameter set $\Theta$, this parameter set being implicit in the density itself; henceforth the dependency on the parameter $\Theta$ needs to be explicitly written as $\pi(\cdot|\Theta)$.  We first derive the solutions for the non-regressed case.  Letting $\widehat\Theta$ denote the current estimate of $\Theta$, we iteratively find
\begin{equation}
Q(\Theta,\widehat{\Theta})=\mathbb{E}\left( \log(L^c)|{\cal X}_1,\ldots,{\cal X}_n,\widehat\Theta \right),
\label{Q1}
\end{equation}
(E step) where $L^c$ is the complete-data likelihood, as given in (\ref{completeLik}), 
and find $\Theta^*$, where $Q(\Theta^*,\widehat\Theta)\geq Q(\widehat\Theta,\widehat\Theta)$ (M step), subsequently setting $\widehat\Theta=\Theta^*$.  Now  $Q(\Theta,\widehat\Theta)$ can, with some algebra (see Appendix B), be written in the tractable form
\begin{flalign}\nonumber
&Q(\Theta,\widehat\Theta)= &\\ \nonumber
&\sum_{i=1}^n \sum_{\ell_1=1}^K\cdots\sum_{\ell_{T_i}=1}^K \left[ \log(\alpha_{\ell_1})+\sum_{t=2}^{T_i}\log(\beta_{\ell_{t-1}\ell_t})\right]\mathbb{P}(Z_{i1}=\ell_1|{\cal X}_i,\widehat\Theta)\prod_{s=2}^{T_i}\mathbb{P}(Z_{is}=\ell_s|Z_{i(s-1)}=\ell_{s-1},{\cal X}_i,\widehat\Theta) &\\ \nonumber
&+ \sum_{i=1}^n \sum_{\ell_1=1}^K\cdots\sum_{\ell_{T_i}=1}^K \left[ \log(\pi({\bf X}_{i1}|Z_{i1}=\ell_1,\Theta))+\sum_{t=2}^{T_i}\log(\pi({\bf X}_{it}|{\bf X}_{i(t-1)},Z_{it}=\ell_t,\Theta))\right] &\\
&\hspace{9pc}\cdot \mathbb{P}(Z_{i1}=\ell_1|{\cal X}_i,\widehat\Theta)\prod_{s=2}^{T_i}\mathbb{P}(Z_{is}=\ell_s|Z_{i(s-1)}=\ell_{s-1},{\cal X}_i,\widehat\Theta).&
\end{flalign}

In computing a valid $\Theta^*$ which increases the value of $Q(\widehat\Theta,\widehat\Theta)$, we wish the number of terms to be summed to grow linearly with respect to time, and not exponentially.  To this end, we utilize recursive relationships to compute the conditional distributions $\mathbb{P}(Z_{it}|Z_{i(t-1)},{\cal X}_i,\widehat\Theta)$ and the marginal distributions $\mathbb{P}(Z_{it}|{\cal X}_i,\widehat\Theta)$.  Again leaving off the subscript $i$ where obvious, it can be shown (see Appendix \ref{AppendZdists}) that for $2\leq t\leq T$,
\begin{equation}
\mathbb{P}(Z_t|Z_{t-1},{\cal X},\Theta) \propto q(Z_t|Z_{t-1}),
\end{equation}
where
\begin{equation}
q(Z_{T}|Z_{T-1})=\beta_{Z_{T-1}Z_{T}}\pi({\bf X}_{T}|{\bf X}_{T-1},Z_{T},\Theta),
\label{eq1}
\end{equation}
and, for $2\leq t< T$,
\begin{equation}
q(Z_t|Z_{t-1}) = \beta_{Z_{t-1}Z_{t}}\pi({\bf X}_{t}|{\bf X}_{t-1},Z_{t},\Theta)\sum_{\ell_{t+1}=1}^Kq(Z_{t+1}=\ell_{t+1}|Z_t).
\label{eq2}
\end{equation}
We then need only normalize to obtain the conditional distributions.  To obtain the marginals, we start by computing and normalizing
\begin{equation}
\mathbb{P}(Z_1|{\cal X},\Theta) \propto \alpha_{Z_1}\pi({\bf X}_{1}|Z_1,\Theta)\sum_{\ell_2=1}^Kq(Z_2=\ell_2|Z_1),
\label{eq3}
\end{equation}
and recursively obtain
\begin{equation}
\mathbb{P}(Z_t|{\cal X},\Theta) = \sum_{\ell_{t-1}=1}^K\mathbb{P}(Z_{t-1}=\ell_{t-1}|{\cal X},\Theta)\mathbb{P}(Z_t|Z_{t-1}=\ell_{t-1},{\cal X},\Theta).
\end{equation}
This can be derived in a similar fashion as that found in Appendix \ref{AppendZdists}.

Using the method of Lagrange multipliers, we find that the value of $\alpha_k$ that maximizes $Q(\Theta,\widehat\Theta)$ is
\begin{equation}
\alpha_k^*=\frac{1}{n}\sum_{i=1}^n\mathbb{P}(Z_{i1}=k|{\cal X}_i,\widehat\Theta),
\end{equation}
and similarly the optimizing value of $\beta_{hk}$ is
\begin{equation}
\beta_{hk}^*=\frac{\sum_{i=1}^n\sum_{t=2}^{T_i}\mathbb{P}(Z_{i(t-1)}=h|{\cal X}_i,\widehat\Theta)\mathbb{P}(Z_{it}=k|Z_{i(t-1)}=h,{\cal X}_i,\widehat\Theta)}{\sum_{i=1}^n\sum_{t=2}^{T_i}\mathbb{P}(Z_{i(t-1)}=h|{\cal X}_i,\widehat\Theta)}.
\end{equation}
To update $\lambda$, $\boldsymbol\mu_k$ and $\Sigma_k$, $k=1,\ldots,K$, we employ a coordinate ascent method.  We initialize $\lambda^*=\hat\lambda$, $\boldsymbol\mu_k^*=\hat{\boldsymbol\mu}_k$, and $\Sigma_k^*=\widehat\Sigma_k$, and then find
\begin{eqnarray}\nonumber
\lambda^*&=&\argmax{\lambda}Q(\{\lambda,\Theta^*\setminus\{\lambda^*\}\},\widehat\Theta) \\
&=&\frac{\sum_{i=1}^n\sum_{t=2}^{T_i}\sum_{\ell_{t}=1}^K\mathbb{P}(Z_{it}=\ell_t|{\cal X}_i,\widehat\Theta)({\bf X}_{it}-{\bf X}_{i(t-1)})'\Sigma_k^{-1}(\boldsymbol\mu_k-{\bf X}_{i(t-1)})}{\sum_{i=1}^n\sum_{t=2}^{T_i}\sum_{\ell_{t}=1}^K\mathbb{P}(Z_{it}=\ell_t|{\cal X}_i,\widehat\Theta)(\boldsymbol\mu_k-{\bf X}_{i(t-1)})'\Sigma_k^{-1}(\boldsymbol\mu_k-{\bf X}_{i(t-1)})},
\end{eqnarray}
\begin{flalign}\nonumber
&\boldsymbol\mu_k^*=\argmax{\boldsymbol\mu_k}Q(\{\boldsymbol\mu_k,\Theta^*\setminus\{\boldsymbol\mu_k^*\}\},\widehat\Theta) &\\
&\hspace{1.2pc}=\frac{  \sum_{i=1}^n \big\{  \mathbb{P}(Z_{i1}=k|{\cal X}_i,\widehat\Theta){\bf X}_{i1}+\lambda \sum_{t=2}^{T_i}\mathbb{P}(Z_{it}=k|{\cal X}_i,\widehat\Theta)({\bf X}_{it}-(1-\lambda){\bf X}_{i(t-1)}) \big\}}{\sum_{i=1}^n \big\{  \mathbb{P}(Z_{i1}=k|{\cal X}_i,\widehat\Theta) +\lambda^2\sum_{t=2}^{T_i}\mathbb{P}(Z_{it}=k|{\cal X}_i,\widehat\Theta) \big\}},&
\end{flalign}
\begin{flalign}\nonumber
&\Sigma_k^*=\argmax{\Sigma_k}Q(\{\Sigma_k,\Theta^*\setminus\{\Sigma_k^*\}\},\widehat\Theta) &\\
&\hspace{1.2pc}=\frac{\sum_{i=1}^n \big\{  \mathbb{P}(Z_{i1}=k|{\cal X}_i,\widehat\Theta)({\bf X}_{i1}-\boldsymbol\mu_k)({\bf X}_{i1}-\boldsymbol\mu_k)'    +\sum_{t=2}^{T_i}\mathbb{P}(Z_{it}=k|{\cal X}_i,\widehat\Theta){\bf H}_{it}{\bf H}_{it}' \big\}}{\sum_{i=1}^n \big\{  \mathbb{P}(Z_{i1}=k|{\cal X}_i,\widehat\Theta) +\sum_{t=2}^{T_i}\mathbb{P}(Z_{it}=k|{\cal X}_i,\widehat\Theta) \big\}}&
\end{flalign}
for $k=1,\ldots,K$, where ${\bf H}_{it}={\bf X}_{it}-\lambda\boldsymbol\mu_k-(1-\lambda){\bf X}_{i(t-1)}$.  See Appendix D for proofs of these solutions.

In the case of regressed cluster probabilities, the complete-data likelihood in $Q(\Theta,\widehat{\Theta})$ is that found in (\ref{completeLikReg}).  To obtain the correct updates for $\lambda$, $\boldsymbol\mu_k$, and $\Sigma_k$, $k=1,\ldots,K$, simply replace $\mathbb{P}(Z_{it}|{\cal X}_{i},\Theta)$ above with $\mathbb{P}(Z_{it}|{\cal X}_{i},\Theta,{\bf w}_{i1},\ldots,{\bf w}_{iT_i})$; these latter distributions are easily computed by first replacing $\alpha_k$ with $\alpha_k({\bf w}_{i1})$ in (\ref{eq3}) and $\beta_{hk}$ with $\beta_{hk}({\bf w}_{it})$ in  (\ref{eq1}) and (\ref{eq2}), and then proceeding with the recursive relationships previously outlined.  To obtain updates for $\delta_{0k}$, $\boldsymbol\gamma_{0k}$, $\delta_{hk}$, and $\boldsymbol\gamma_{hk}$, $h=1,\ldots,K$, $k=1,\ldots,K-1$, one can use a numerical optimization method (we implemented the quasi-Newton BFGS method) to find
\begin{equation}
\argmax{\{\delta_{0k},\boldsymbol\gamma_{0k}:k=1,\ldots,K\}}=\sum_{i=1}^n\sum_{k=1}^K\mathbb{P}(Z_{i1}=k|{\cal X}_i,\widehat\Theta)\log(\alpha_k({\bf w}_{i1}))
\label{bfgs1}
\end{equation}
\begin{equation}
\argmax{\{\delta_{hk},\boldsymbol\gamma_{hk}:k=1,\ldots,K\}}=\sum_{i=1}^n\sum_{t=2}^{T_i}\sum_{k=1}^K\mathbb{P}(Z_{i(t-1)}=h|{\cal X}_i,\widehat\Theta) \mathbb{P}(Z_{it}=k|Z_{i(t-1)}=h,{\cal X}_i,\widehat\Theta)\log(\beta_{hk}({\bf w}_{it}))
\label{bfgs2}
\end{equation}
for $h=1,\ldots,K$.  Since (\ref{bfgs1}) and (\ref{bfgs2}) are both concave, finding the global maximum  at each M-step is assured upon convergence of the BFGS algorithm.

\section{Simulation Study}
We simulated two illustrative data sets, one without and one with regressed cluster probabilities, where $p$, the dimension of each ${\bf X}_{it}$, is two.  We then simulated 100 data sets where $p=5$, 50 without and 50 with regressed cluster probabilities (thus a total of 102 simulated data sets).  The resulting clustering was compared in the variation of information (VI) (Meil{\u{a}} (2003)) and in the corrected Rand index (CRI) (Hubert and Arabic (1985)).  The VI, a true metric on clusterings, yields a value of 0 for identical clusterings; higher values indicate more disparate clustering assignments.  The corrected Rand index, adjusted to account for chance, yields a value of 1 for perfect cluster agreement and has an expected value of 0 for random cluster assignments.  The results are both in the text as well as in Table \ref{simResultsTable}.

Each simulated data set (when $p=2$ and when $p=5$) without regressed cluster probabilities was achieved according to the following.  The number of objects $n$ was 100, the maximum number of time steps $\max_i\{T_i\}$ was 10, and the number of clusters $K$ was 5.  To check whether our method can accurately group the objects, the simulations consisted of a wide range of possible parameter values.  We obtained the simulated data in the following way.  The quantities $T_i$ were uniformly sampled from 1 to 10;  $\lambda$ was drawn from a beta distribution with both parameters equal to 10 (the mean of the distribution is 0.5);  $\boldsymbol\alpha$ was drawn from a Dirichlet distribution with parameters all equalling 10;  $\boldsymbol\mu_k$'s were drawn from a multivariate normal distribution with mean ${\bf 0}$ and diagonal covariance matrix with diagonal entries equal to 20;  $\Sigma_k$'s were drawn from an inverse Wishart distribution with $p+4$ degrees of freedom and diagonal scale matrix whose diagonal entries were 3.  The $h^{th}$ row of $\beta$ was set to be proportional to
\begin{equation}\nonumber
\left(\frac{1}{\|\boldsymbol\mu_1 -\boldsymbol\mu_h\|},\ldots,\frac{1}{\|\boldsymbol\mu_{h-1} -\boldsymbol\mu_h\|},15\times \max_{k\neq h}\left\{\frac{1}{\|\boldsymbol\mu_k -\boldsymbol\mu_h\|}\right\},\frac{1}{\|\boldsymbol\mu_{h+1} -\boldsymbol\mu_h\|},\ldots,\frac{1}{\|\boldsymbol\mu_K -\boldsymbol\mu_h\|} \right).
\end{equation}
This formulation of $\beta$ was chosen to put most of the probability on remaining in the same cluster and to force the probability of jumping to a different cluster to decrease as the distance to that different cluster increased.
The $Z_{it}$'s were then generated according to $\boldsymbol\alpha$ and $\beta$, and the ${\cal X}_i$'s were generated according to (\ref{transition1}) and (\ref{transition2}).

\begin{figure}
\centering
\subfloat[Initial Clustering]{
\includegraphics[width=0.52\textwidth]{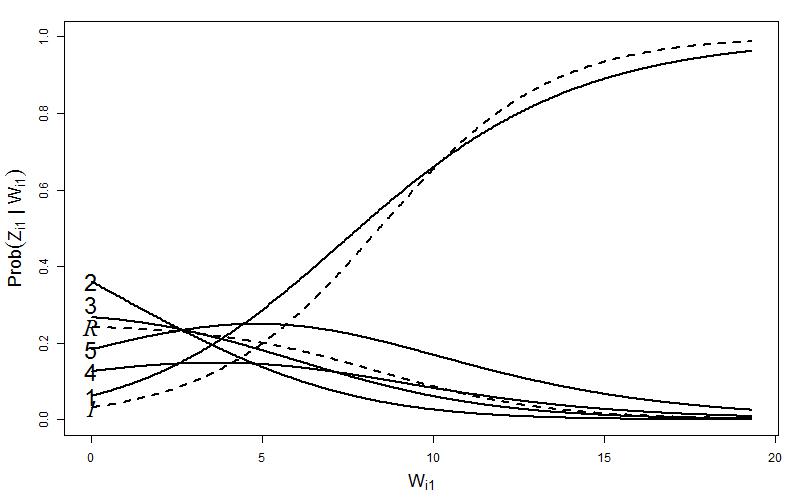}
}
\subfloat[From Cluster 1]{
\includegraphics[width=0.52\textwidth]{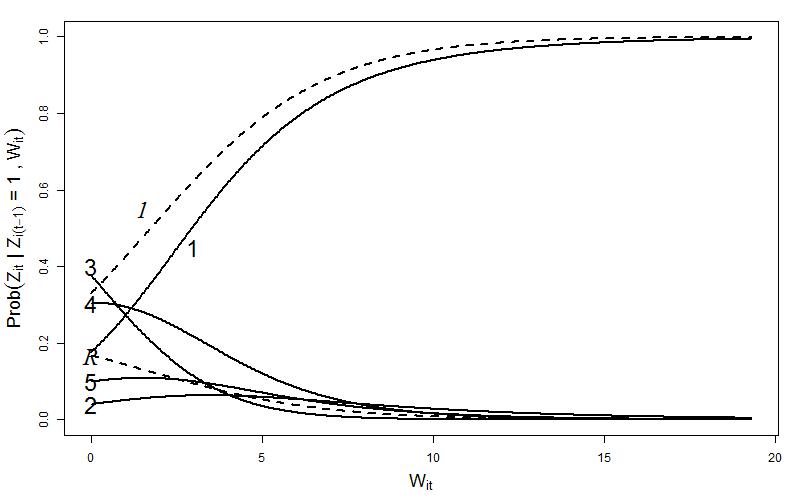}
}\\
\subfloat[From Cluster 2]{
\includegraphics[width=0.52\textwidth]{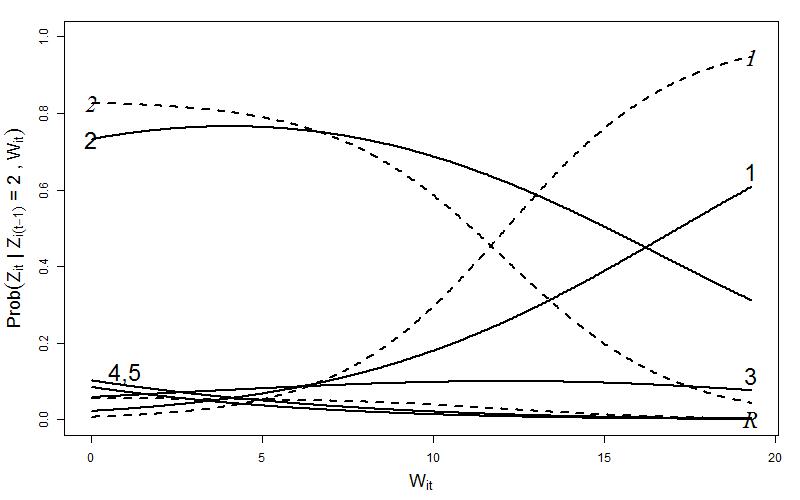}
}
\subfloat[From Cluster 3]{
\includegraphics[width=0.52\textwidth]{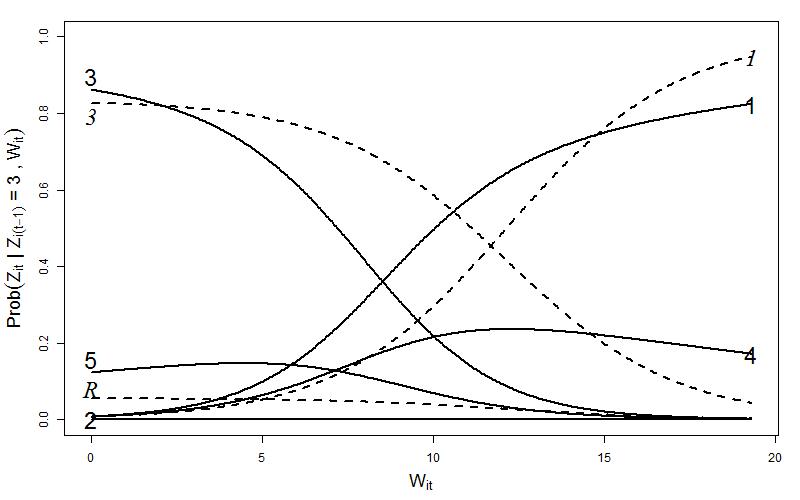}
}\\
\subfloat[From Cluster 4]{
\includegraphics[width=0.52\textwidth]{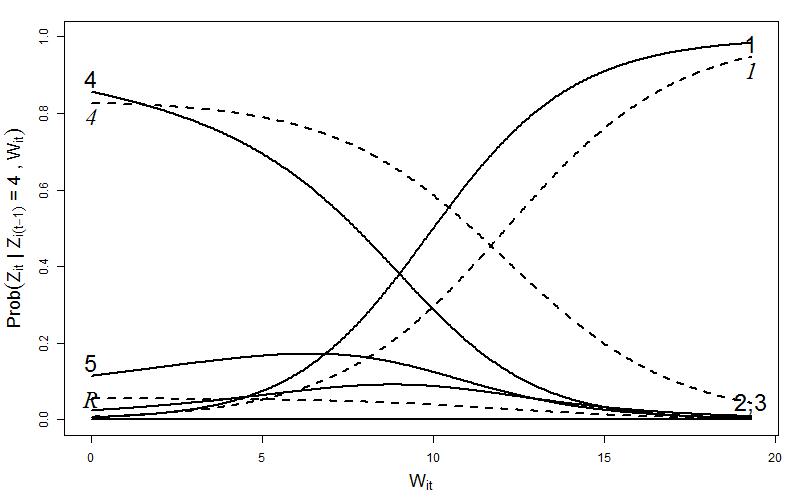}
}
\subfloat[From Cluster 5]{
\includegraphics[width=0.52\textwidth]{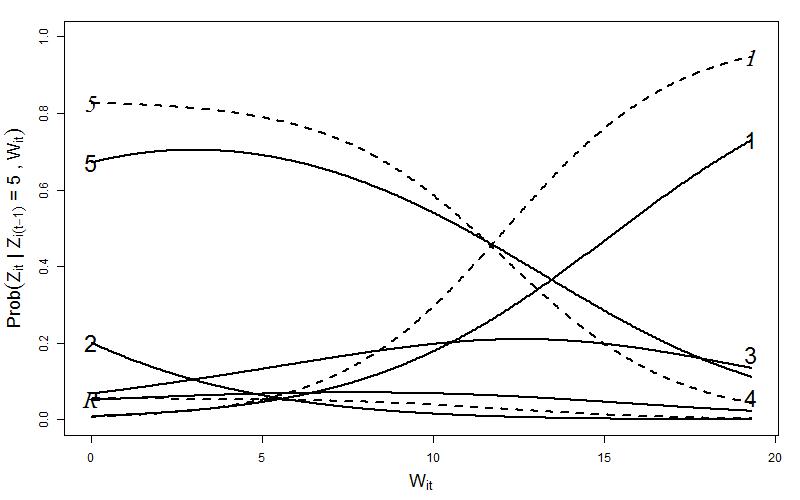}
}
\caption{\footnotesize Cluster probabilities for illustrative simulated data with regressed cluster probabilities. Solid lines are estimated probability curves, dashed lines are true curves. Each curve represents, for the varying values of the explanatory variable, the probability of belonging to the cluster whose number is adjacent to the curve. There are only two or three (depending on the figure) unique true probability curves which are labeled in italics, hence ``R" indicates the probability curve corresponding to those remaining cluster numbers which are not unique, e.g., clusters 2 to 5 in the initial clustering plot. \normalsize}
\label{sim_multLogit}
\end{figure}

The simulations with regressed cluster probabilities were generated in a similar manner, but we allowed $\boldsymbol\alpha$ and $\boldsymbol\beta$ to depend on a positive valued covariate ($w_{it}>0$) which affects the probability of belonging to cluster one.  A high value of the covariate implied a tendency to belong to cluster one while a low value of the covariate implied a tendency to avoid belonging to cluster one.  Beyond this there was no discriminating influence from the covariate on the probabilities of belonging or transitioning to the other clusters.  Just as in the simulation without a covariate, there was also an inherent tendency to remain in the same cluster rather than switch (though this could, for objects not in cluster one, potentially be countered by extremely large covariate values).  This was accomplished in the following way.  For $i=1,\ldots,n$, the explanatory variable $w_{i1}$ ($d=1$) was drawn from a chi-squared distribution with five degrees of freedom, and for $t=2,\ldots,T_i$, $w_{it}$ was drawn from a normal distribution with mean $w_{i(t-1)}$ and standard deviation 0.5.  In this way the objects' covariate values had a certain degree of smoothness over time.  The parameters $\gamma_{0k}$ and, for $h=1,\ldots,K$, $\gamma_{hk}$ were set to be $\log(1.5)$ if $k= 1$, and 0 otherwise.  We set $\delta_{0k}$ and, for $h=1,\ldots,K$, $\delta_{hk}$ equal to $-5\log(1.5)$ if $k= 1$, and 0 otherwise; in addition, $\log(15)$ was added to $\delta_{hh}$, $h<K$, and $-\log(15)$ was added to $\delta_{Kk}$, $k<K-1$.  This leads to the situation described above where there is an increased chance of staying in the same cluster than switching to another cluster; further, if $w_{it}$ is 5 (the mean of $w_{it}$) there is no particular tendency or avoidance of belonging to cluster one, if $w_{it}>5$ there is an increased probability of belonging to cluster one, and if $w_{it}<5$ there is a decreased probability of belonging to cluster one.  The solid line curves in Figure \ref{sim_multLogit} graphically demonstrate the initial cluster probabilities and transitional cluster probabilities for varying values of the covariate $w_{it}$.
\begin{table}[htb]
\centering
\begin{tabular}{|c|c|c||c|c|}
\hline
&\multicolumn{2}{c||}{Without regressed cluster probabilities}&\multicolumn{2}{c|}{With regressed cluster probabilities}\\ \cline{2-5}
&Proposed method&k-means&Proposed method&k-means\\ \hline
VI&0.0180 (0)&0.5516 (0.5175)&0.0531 (0)&0.7413 (0.7429)\\ \hline
CRI&0.9953 (1)&0.8426 (0.8609)&0.9780 (1)&0.7729 (0.7953)\\ \hline
\end{tabular}
\caption{From the simulations run without and with regressed cluster probabilities, the estimated clustering results are compared to the true cluster assignments for both k-means and our proposed approach using both variation of information (VI) and the corrected Rand index (CRI).  Means over 50 simulations are given with medians in parentheses.}
\label{simResultsTable}
\end{table}

\begin{figure}[htb]
\centering
\subfloat[Without regressed cluster probabilities]{
\includegraphics[height=0.5\textwidth,width=0.5\textwidth]{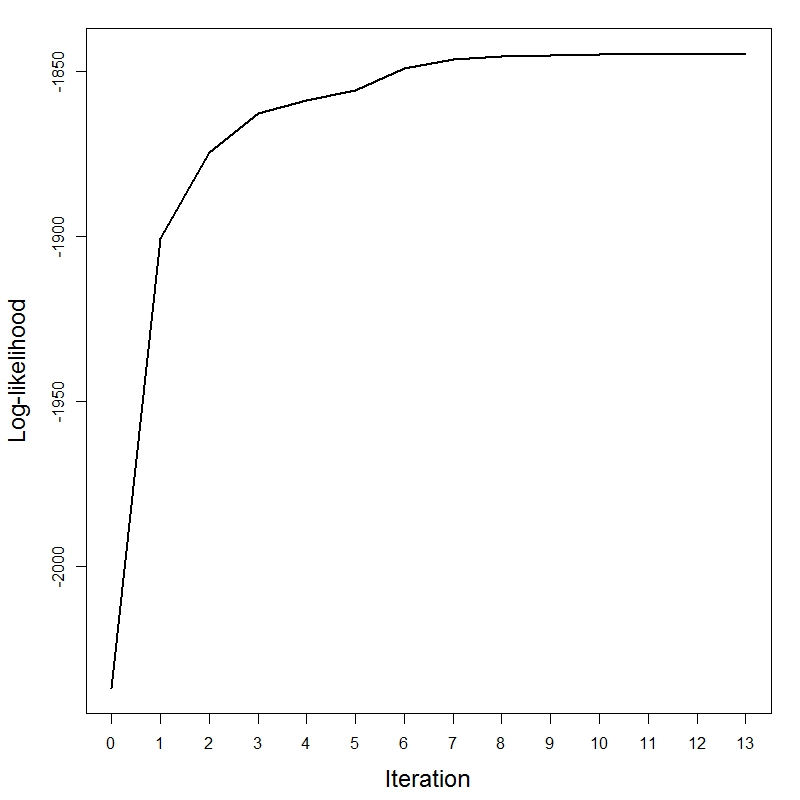}
\label{noRT_llik}
}
\subfloat[With regressed cluster probabilities]{
\includegraphics[height=0.5\textwidth,width=0.5\textwidth]{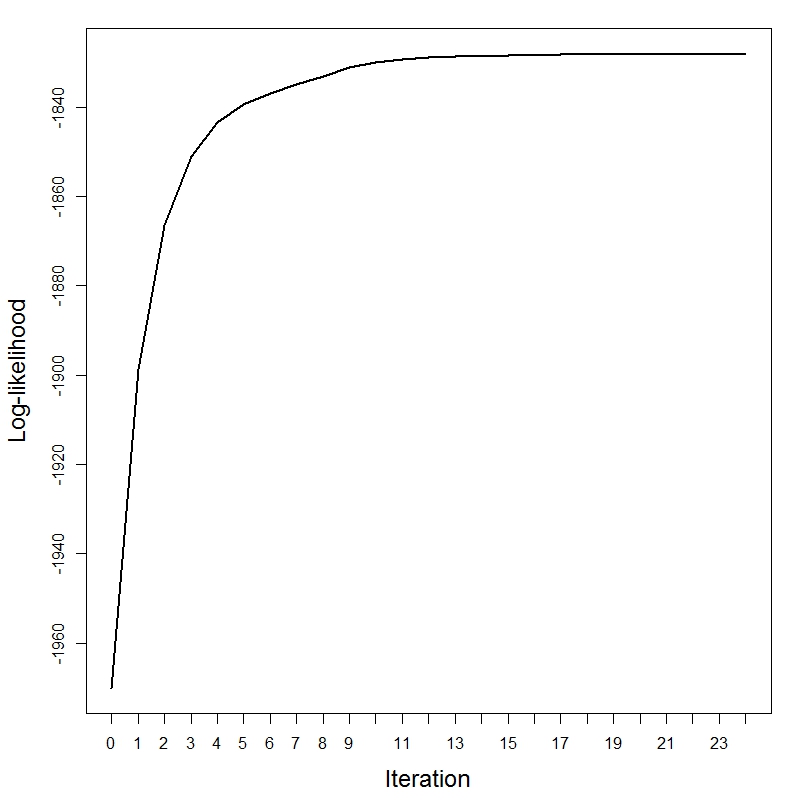}
\label{RT_llik}
}
\caption{Log-likelihood over the iterations for the two illustrative simulations.}
\end{figure}

\begin{figure}
\centering
\includegraphics[height=17cm,width=17cm]{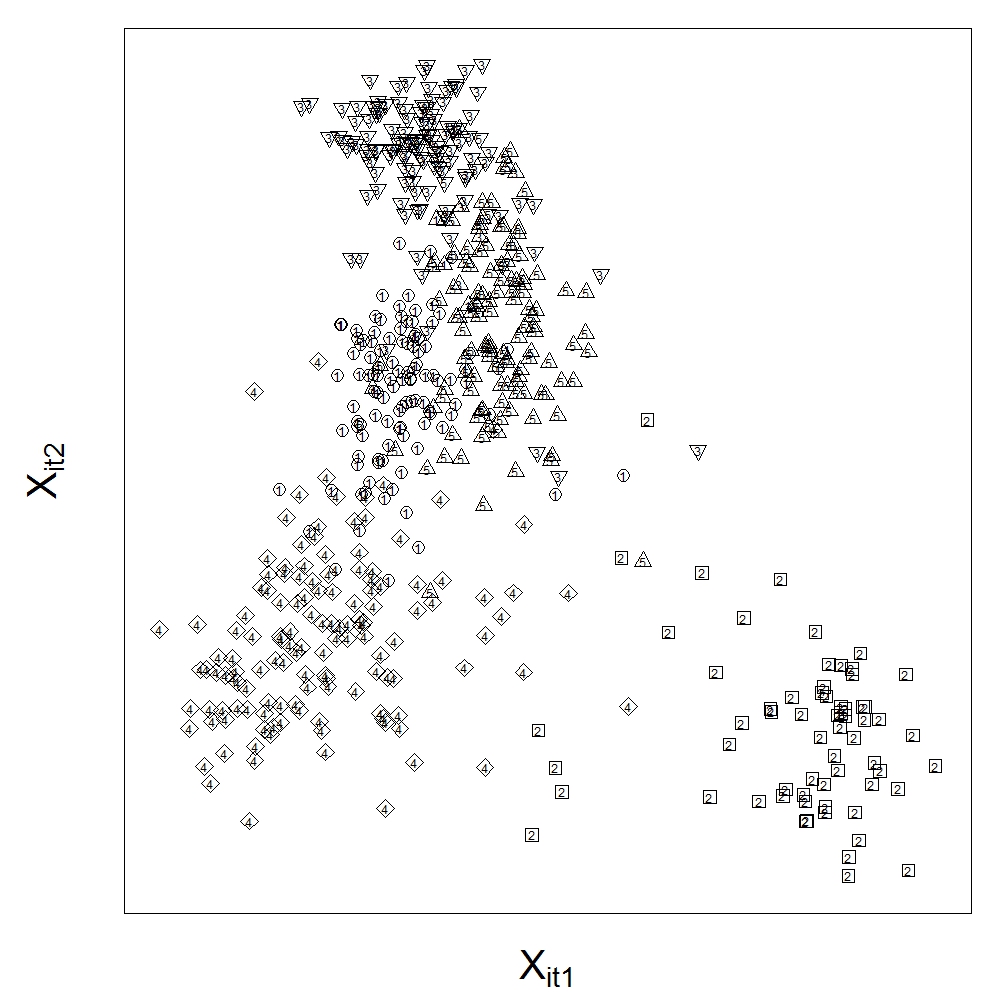}
\caption{Plot of simulated data without regressed cluster probabilities, where the horizontal and vertical axes correspond to the two variables on which clustering is performed.  Numbers correspond to true clustering, shapes correspond to estimated hard clustering assignments.}
\label{noRT_2dplot}
\end{figure}

For the illustrative simulation without regressed cluster probabilities ($p=2$), the VI from using the proposed approach was 0.2679 and CRI was 0.9232.  Simply using k-means and choosing the best k-means result from 15 starting positions (which was used to initialize the generalized EM algorithm), led to a VI of 1.031 and CRI of 0.6615.  Figure \ref{noRT_llik} gives the plot of the log-likelihood over the iterations of the generalized EM algorithm of the proposed method, starting with the initialized values.  We see that the algorithm converged within a small number of iterations.  Figure \ref{noRT_2dplot} shows the simulated data, where the numbers correspond to the true cluster assignments, and the shapes correspond to the estimated hard cluster assignments.  Regarding the 50 simulations without regressed cluster probabilities ($p=5$), the mean (median) VI was 0.0180 (0) and CRI was 0.9953 (1).  Using k-means, the mean (median) VI was 0.5516 (0.5175) and CRI was 0.8426 (0.8609).  Using a UNIX machine with a 2.40 GHz processor, the mean (median) elapsed time for our proposed method was 0.5507 sec (0.5120 sec).  These values of VI and CRI imply that our model performs extremely well, and much better than the na\"{i}ve k-means approach.

\begin{figure}
\centering
\includegraphics[height=17cm,width=17cm]{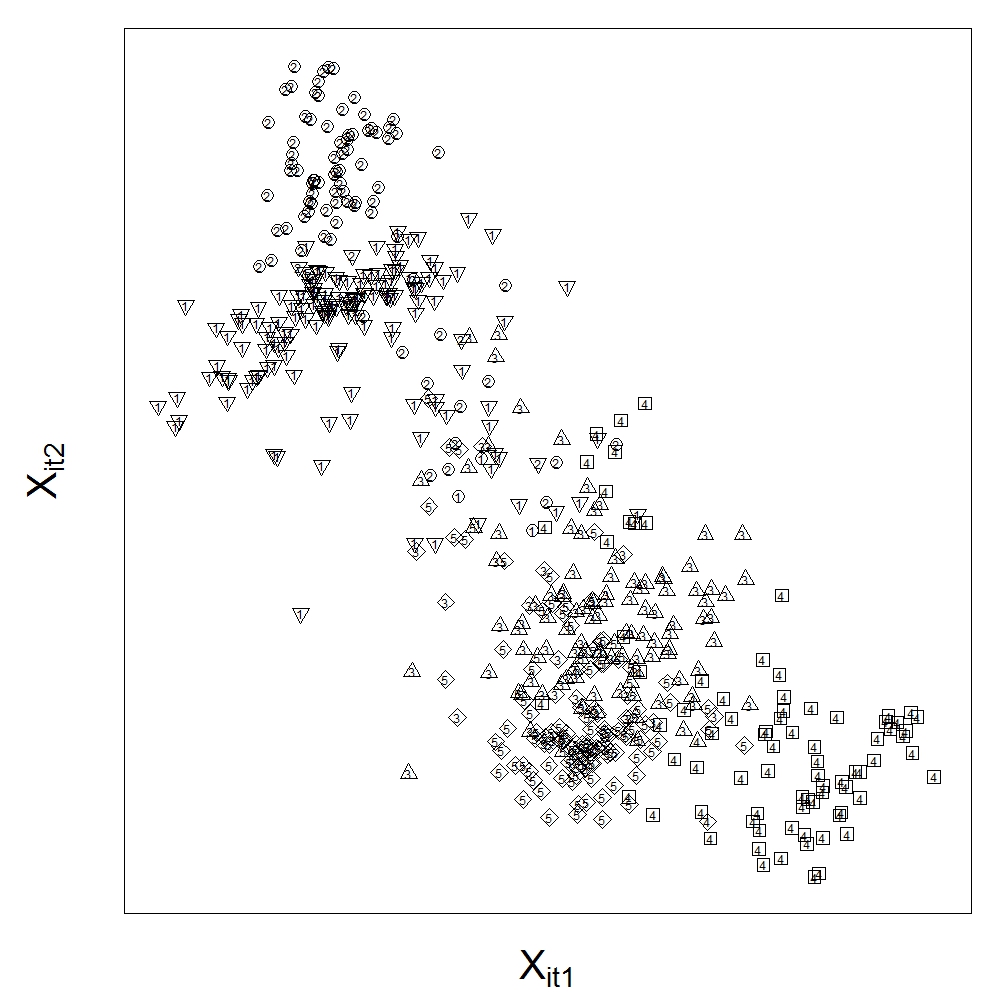}
\caption{Plot of simulated data with regressed cluster probabilities, where the horizontal and vertical axes correspond to the two variables on which clustering is performed.  Numbers correspond to true clustering, shapes correspond to estimated hard clustering assignments.}
\label{RT_2dplot}
\end{figure}

For the illustrative simulation with regressed cluster probabilities ($p=2$), the VI from using the proposed approach was 0.5080 and the CRI was 0.8232.  Simply using k-means resulted in a VI of 1.342 and a CRI of 0.5147.  Figure \ref{RT_llik} gives the plot of the log-likelihood over the iterations of the generalized EM algorithm in the proposed method, starting with the initialized values.  From this we again see that the algorithm reached convergence within a small number of iterations.  Figure \ref{RT_2dplot} shows the simulated data, where the numbers correspond to the true cluster assignments, and the shapes correspond to the estimated hard cluster assignments.  Regarding the 50 simulations with regressed cluster probabilities ($p=5$), the mean (median) VI was 0.0531 (0) and CRI was 0.9780 (1).  Using k-means, the mean (median) VI was 0.7413 (0.7429) and CRI was 0.7729 (0.7953).  The mean (median) elapsed time for our proposed method was 41.82 sec (35.41 sec).  As in the case for non-regressed cluster probabilities, these values of VI and CRI imply that our model performs extremely well, and much better than the na\"{i}ve k-means approach.  

We can also investigate whether the model captured the effect of the explanatory variable.  An effective visualization of this is to plot the cluster probabilities $\alpha_k(w_{i1})$ and $\beta_{hk}(w_{it})$ for $h,k=1,\ldots,K$, over the range of $w_{it}$.  This is more intuitive than trying to look at the individual $\delta_{hk}$'s and $\gamma_{hk}$'s, allowing the user to see exactly how the explanatory variables affect the cluster probabilities.  Figure \ref{sim_multLogit} gives these plots, where each line corresponds to probabilities of belonging to a particular cluster.  Here we can see that the estimated probability curves (solid lines) correspond very closely to the true probability curves (dashed lines), implying that the estimation method captured the effect of the explanatory variable effectively.

\vspace{-1pc}
\section{U.S. Congressional Data}
We collected 15 variables measuring legislative activity for Democrats in the House of Representatives in the 101$^{st}$-110$^{th}$ Congresses (1989-2008).  Because there are multiple variables for a single concept, we combined these indicators into eight indices --- paying attention to the home district; showboating (e.g., making speeches and writing editorials); voting with the party on roll calls; giving campaign funds to the party; specializing in particular policy issues; building bipartisan coalitions; overall fundraising; and lawmaking (e.g., introducing legislation).  See Appendix \ref{AppendMC} for more details on these variables.  Thus ${\bf X}_{it}$ is the eight-dimensional vector corresponding to the scores on these eight indices measured for the $i^{th}$ member of Congress (MC) at his or her $t^{th}$ term.  There were $n=539$ unique MC's that served one or more terms over the 20 years included in the study.

We wished to determine the impact of the MCs' ideology on the MCs' behaviors.  To this end we included a covariate for MC ideology to help predict initial cluster assignments for individual MCs, as well as transitions between clusters across time. Our measure of ideology was the common-space NOMINATE score. These scores are based on a multidimensional scaling algorithm for roll call voting developed by Poole and Rosenthal (1997) and are a function of how often each individual MC makes the same vote choice as each other MC. The algorithm allows all legislators to be arrayed on a scale of about $-1$ to $1$, with negative scores indicating liberal ideology and positive scores indicating conservative ideology. Since the sample is solely Democrats, the scores are skewed toward the negative end of the scale, ranging from $-0.725$ to $0.190$.  In addition to this ideology variable, we also included time dummy variables in order to account for behavioral shifts across the entire party that occur during the various Congresses (each Congress has its own dummy variable).  Lastly, for the 101$^{st}$ Congress we differentiated between true freshmen politicians as well as those MCs who have served previous terms not included in the data by allowing the true freshmen (for all Congresses) and the first observed MCs (of the 101$^{st}$ Congress) to have different initial clustering coefficients $\delta_{0k}$ and $\boldsymbol\gamma_{0k}$.  This was accomplished by letting ${\bf w}_{i1}=(ideology,ideology\cdot 1_{\left\{\mbox{First Observed}\right\}},1_{\{t_1<101\}}, 1_{\{t_1=101\}},\ldots,1_{\{t_1=109\}})$, where $ideology$ is the NOMINATE score, $1_{\left\{\mbox{First Observed}\right\}}$ is 1 if not a true freshman and 0 otherwise, $1_{\{t_1<101\}}$ if the MC's first term was before the $101^{st}$ Congress and $1_{\{t_1=s\}}$ is 1 if the $i^{th}$ MC's first term (of the study) was served during the $s^{th}$ Congress and 0 otherwise, for $s=101,102,\ldots,109$.

To determine the number of clusters we used the average silhouette statistic (Rousseeuw (1987)).  We chose this rather than a statistic that is dependent on the number of model parameters, such as AIC or BIC, in order to avoid the number of covariates determining the number of clusters.  That is, the number of clusters ought not to depend on the number of covariates used to explain the clustering.  The results indicated four clusters of legislator types.

\begin{figure}[htb]
\centering
\includegraphics[height=10.cm,width=10.cm]{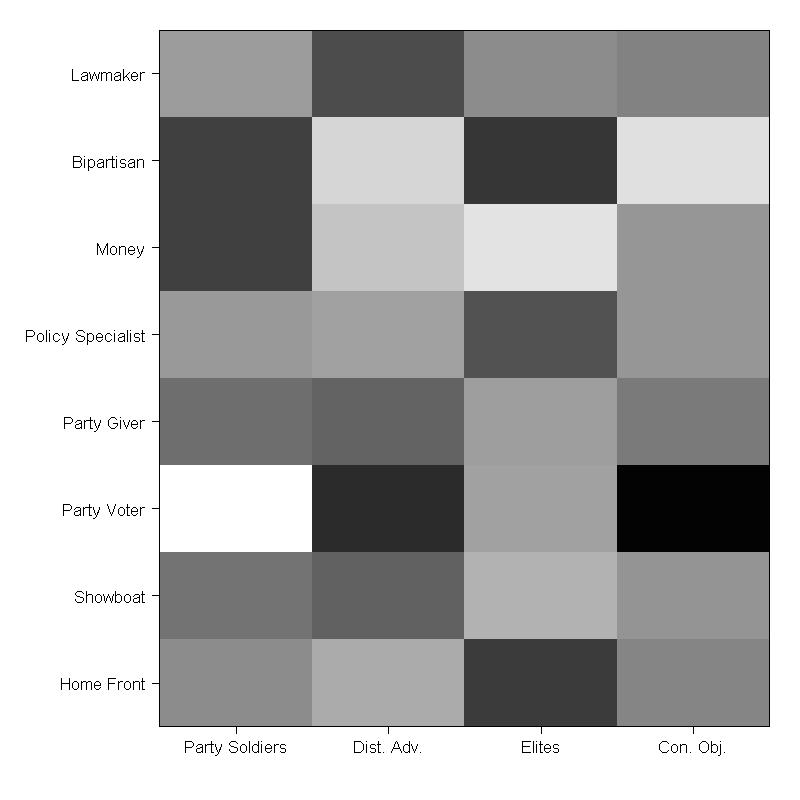}
\caption{\small Cluster means for the Democrat MCs.  The vertical axis corresponds to the behavioral variables on which the MCs were clustered, and the horizontal axis corresponds to the clusters.  Lighter hues indicate higher values, darker hues indicate lower values.}
\label{DemMean}
\end{figure}

We applied our proposed method and obtained the following results.  Figure \ref{DemMean} gives the cluster means $\boldsymbol\mu_k$.  Cluster one, the largest cluster (50\% of $Z_{it}$'s equal 1), consists of MCs that we call party soldiers. This group is marked by its very high score on party voting, but relatively low scores on lawmaking, showboating, and other activities. These are the rank-and-file legislators who toe the party line, but do not distinguish themselves in other ways.  Cluster two, the second largest cluster comprising 26\% of the observations, represents district advocates --- these are legislators who devote their efforts to their home districts, but are not particularly active in the lawmaking process and are not committed partisans.  Cluster three, with 16\% of the observations, are the elites --- MCs who are publicly visible, strongly support the party both in voting and giving of contributions, and who are very active in the lawmaking process. Across the sample period, the actual party leadership (e.g., Richard Gephardt and Nancy Pelosi) fall into this cluster.  The final cluster, the ``conscientious objectors," is the smallest, with just 8\% of observations. These are MCs who are publicly visible (scoring high on the showboat index), are policy specialists, and are very bipartisan in their coalitions, and also have the lowest mean of the four clusters on party support in voting.  These MCs chart their own courses, pursuing their policy goals with less regard for the party leadership's preferences.

\begin{figure}[htb]
\centering
\includegraphics[height=0.4\textheight]{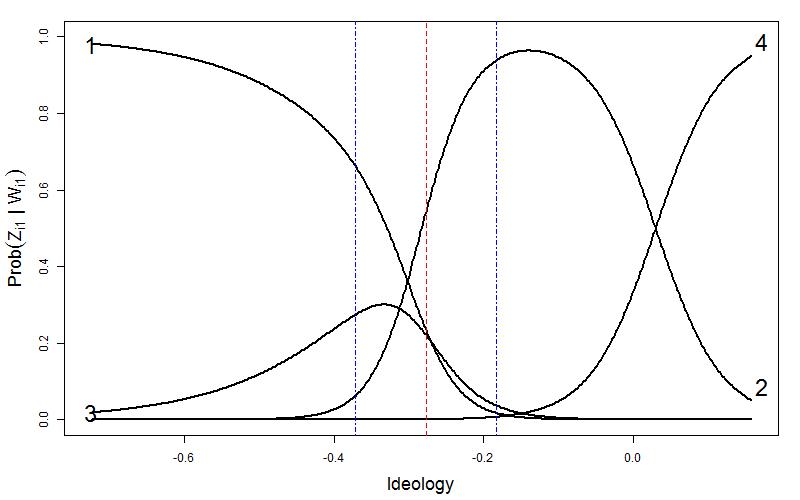}
\caption{\small Initial cluster probabilities for true freshmen in the Congressional data.  Each curve represents, for the varying values along the horizontal axis of the explanatory variable (ideology), the probability of belonging to the cluster whose number is adjacent to the curve.  The horizontal axis spans the range of all MCs ideology;  the dot-dashed lines give the 25\% and 75\% quantiles of ideology, and the long dashed line gives the mean ideology.}
\label{polisciInit}
\end{figure}

The results are in line with intuitions about the relationship between ideology and legislative styles. For example, as shown in Figure \ref{polisciInit}, the probability of falling in the party soldier cluster decreases as the ideology score increases:  strong liberals are more likely than their moderate counterparts to be party soldiers.  On the other hand, the probability of being a conscientious objector increases as the ideology score increases:  moderates are more likely than liberals to fall into this category.  The probabilities associated with belonging to the district advocates or elites clusters are less linear.  The likelihood of being in the elite category peaks just to the left of the mean ideology score --- leaders tend to have more pragmatic outlooks, and, while solidly liberal, are seldom ideologically extreme.  The probability of being a district advocate, on the other hand, is greatest to the right of the mean ideology score.  MCs who focus on their districts often come from heterogeneous constituencies, and, as such, must devote special time and attention to cultivating the district in order to win reelection.

\begin{figure}
\centering
\subfloat[Transition from Party Soldier (cluster 1)]{
\includegraphics[width=0.53\textwidth]{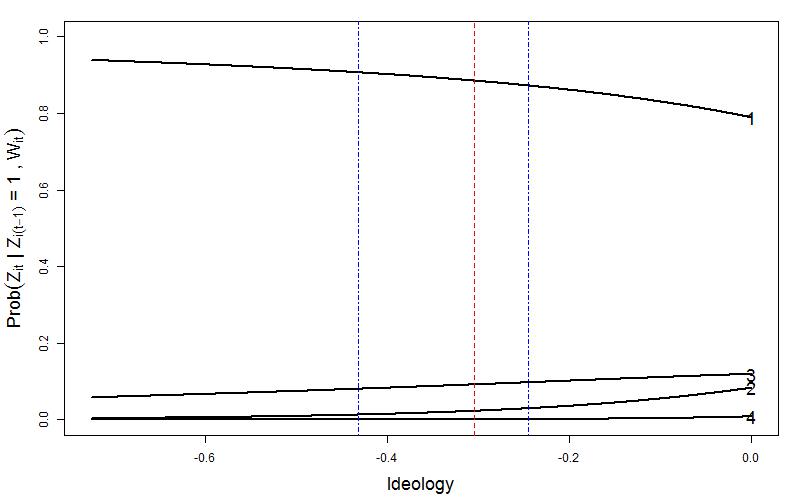}
\label{tr1}
}
\subfloat[Transition from District Advocate (cluster 2)]{
\includegraphics[width=0.53\textwidth]{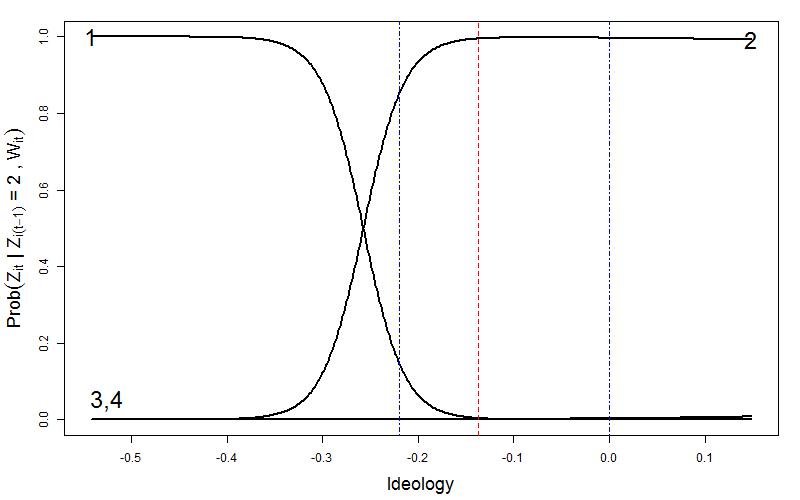}
\label{tr2}
}\\
\subfloat[Transition from Elite (cluster 3)]{
\includegraphics[width=0.53\textwidth]{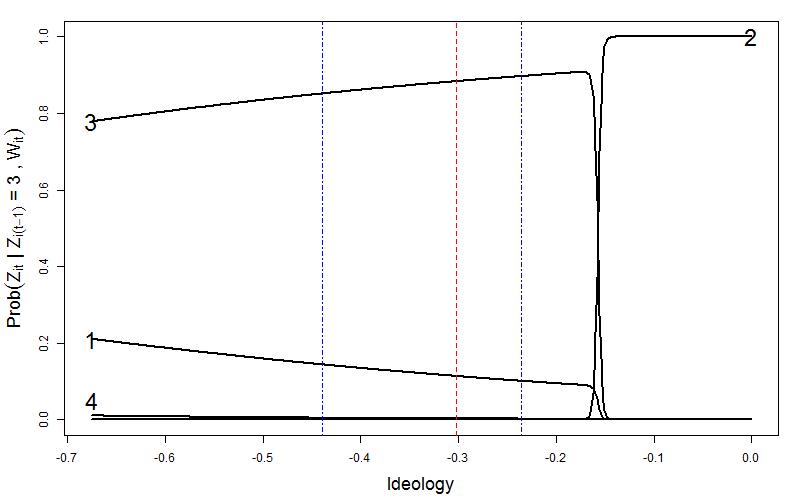}
\label{tr3}
}
\subfloat[Transition from Conscientious Objector (cluster 4)]{
\includegraphics[width=0.53\textwidth]{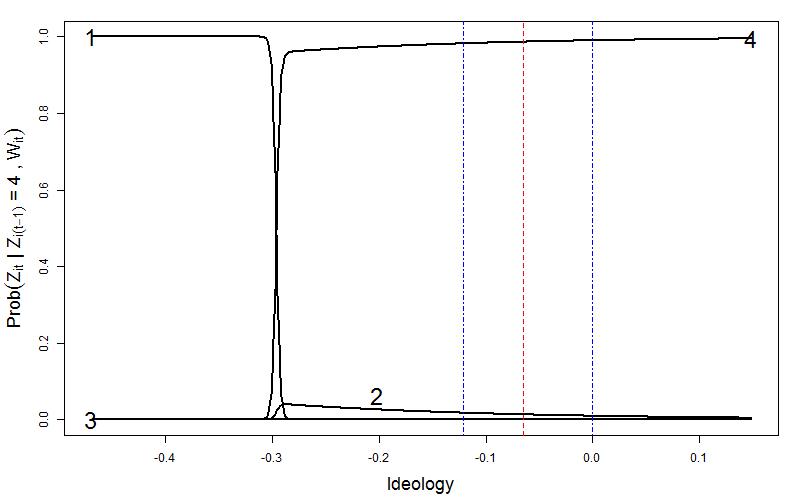}
\label{tr4}
}
\caption{Transition cluster probabilities for Congressional data. Each curve represents, for the varying values along the horizontal axis of the explanatory variable (ideology), the probability of belonging to the cluster whose number is adjacent to the curve given that they currently belong to cluster: (a) 1 (party soldiers) (b) 2 (district advocates) (c) 3 (elites) (d) 4 (conscientious objectors). The horizontal axis spans the range of all MCs ideology; the dot-dashed lines give the 25\% and 75\% quantiles of ideologies of MCs who have belonged to the corresponding cluster, and the long dashed line gives the mean.\normalsize}
\label{polisciProbCurves}
\end{figure}

Figure \ref{polisciProbCurves} demonstrates the relationship between the MCs' ideology and their probability of falling in a particular cluster at time $t + 1$, given their cluster assignment at time $t$. These results largely correspond to intuition as well.  For example, party soldiers at time $t$ are more likely to remain as such at $t + 1$ if they are very liberal.  In contrast, district advocates, conscientious objectors, and elites are the most likely to maintain their cluster assignments across time if they are moderate, although the probability of remaining an elite drops off quickly if the MC becomes too moderate.

\vspace{-1pc}
\section{Discussion}
We developed a clustering model for longitudinal data that accounts for time-varying cluster assignments, time series objects of varying length, and time-dependence via blending cluster effects with individual effects.
We have shown how variables that explain the clustering can be integrated into the model, providing insight into the relationships existing between the clustering and the explanatory variables.  We derived efficient methods for computing the likelihood, as well as efficient estimation methods using the generalized EM algorithm.

While the model given in Section \ref{Regressedclusterprobabilities} is completely identifiable with the constraints that $\delta_{\ell K}=0$ and $\boldsymbol{\gamma}_{\ell K}={\bf 0}$ for $\ell=0,\ldots,K$, the MLE's of the $\delta_{\ell k}$'s and $\boldsymbol{\gamma}_{\ell k}$'s may diverge and/or may not be unique if there are insufficient data points, especially if some cluster transitions are rare.
A simple example of this is has no observed transitions out of, say, cluster $h$.   In such a case, the (possibly non-unique) MLE's $\hat{\delta}_{hh}\rightarrow\infty$, $\hat{\delta}_{hk}\rightarrow-\infty$ for $k\neq h$, and $\hat{\boldsymbol{\gamma}}_{hk}$, $k=1,\ldots,K$, may take any finite values.  As these computational issues are due to the true probability distributions of the cluster assignments not being adequately represented by the data, we would expect that these issues would disappear as $n\rightarrow\infty$.  Future study on the existence and uniqueness of the MLE's would certainly be useful.

As mentioned in Section \ref{Nonregressedclusterprobabilities}, the dependency structure of the observed values ${\bf X}_{it}$ follows a Markov switching model, where at time $t$ all the previous information in ${\cal X}_i$ up to time $t-1$ is captured in ${\bf X}_{i(t-1)}$.  There is, of course, a large number of possible dependency structures that could be represented in the data.  For example, if there was some seasonality of length $s$ in the individual effect, we might change (\ref{Xdepend}) to
\begin{equation}
{\bf X}_{it}|{\bf X}_{i1}\ldots,{\bf X}_{i(t-1)},Z_{i1},\ldots,Z_{iT_i}\stackrel{{\cal D}}{=} {\bf X}_{it}|{\bf X}_{i(t-1)},{\bf X}_{i(t-s)},Z_{it},
\end{equation}
and change (\ref{transition2}) to
\begin{equation}
\pi({\bf X}_{it}|{\bf X}_{i(t-1)},{\bf X}_{i(t-s)},Z_{it}) =N({\bf X}_{it}|(1-\lambda_1-\lambda_s) \boldsymbol{\mu}_{Z_{it}}+\lambda_1{\bf X}_{i(t-1)}+\lambda_s{\bf X}_{i(t-s)},\Sigma_{Z_{it}}).\end{equation}
The computation in such a case should not be drastically different than in the simpler model.  In general, if a more sophisticated Markov switching model is more appropriate to the context, it should be feasible to alter the model accordingly.  See Fr\"uhwirth-Schnatter (2006) for more information on various Markov switching models.

\vspace{-1pc}
\appendix
\vspace{-1pc}
\section*{Appendices}
\vspace{-1pc}
\section{Proof of $\pi(Z_t|Z_{t-1},{\bf X}_1,\ldots,{\bf X}_{t-1})=\beta_{Z_{t-1}Z_t}$}
\label{betaZtZtm1}
We wish to find a tractable form of $\pi(Z_t|Z_{t-1},{\bf X}_1,\ldots,{\bf X}_{t-1})$.  Note that the subscript $i$ is suppressed for ease of notation.  The key equality to show is that $\pi({\bf X}_1,\ldots,{\bf X}_{t-1}|Z_{t-1},Z_t)=\pi({\bf X}_1,\ldots,{\bf X}_{t-1}|Z_{t-1})$.  This is shown by
{\small
\begin{align}\nonumber
&\pi({\bf X}_1,\ldots,{\bf X}_{t-1}|Z_{t-1},Z_t)& \\ \nonumber
&=\sum_{Z_1,\ldots,Z_{t-2}}\pi({\bf X}_1,\ldots,{\bf X}_{t-1}|Z_1,\ldots,Z_t)\pi(Z_1,\ldots,Z_t)/\big( \pi(Z_{t-1})\beta_{Z_{t-1}Z_t} \big)& \\ \nonumber
&=\sum_{Z_1,\ldots,Z_{t-2}} \alpha_{Z_1}\pi({\bf X}_1|Z_1)\beta_{Z_1Z_2}\pi({\bf X}_2|{\bf X}_1,Z_2)
\cdots \beta_{Z_{t-2}Z_{t-1}}\pi({\bf X}_{t-1}|{\bf X}_{t-2},Z_{t-1})/\pi(Z_{t-1})& \\ \nonumber
&=\sum_{Z_1,\ldots,Z_{t-2}}\pi({\bf X}_1,\ldots,{\bf X}_{t-1}|Z_1,\ldots,Z_{t-1})\pi(Z_1,\ldots,Z_{t-1})/\pi(Z_{t-1})& \\ \nonumber
&=\sum_{Z_1,\ldots,Z_{t-2}}\pi({\bf X}_1,\ldots,{\bf X}_{t-1},Z_1,\ldots,Z_{t-1})/\pi(Z_{t-1})& \\ \nonumber
&=\pi({\bf X}_1,\ldots,{\bf X}_{t-1}|Z_{t-1}).&
\end{align}
}
From this it is straightforward to obtain the result in the following way
{\small
\begin{equation*}
\pi(Z_t|Z_{t-1},{\bf X}_1,\ldots,{\bf X}_{t-1})= \frac{\pi({\bf X}_1,\ldots,{\bf X}_{t-1},Z_t|Z_{t-1})}{\pi({\bf X}_1,\ldots,{\bf X}_{t-1}|Z_{t-1})} = \frac{\pi({\bf X}_1,\ldots,{\bf X}_{t-1}|Z_{t-1},Z_t)\beta_{Z_{t-1}Z_t}}{\pi({\bf X}_1,\ldots,{\bf X}_{t-1}|Z_{t-1})} = \beta_{Z_{t-1}Z_t}.
\end{equation*}
}

\vspace{-1pc}
\section{Deriving the Tractable Form of $Q(\Theta,\widehat\Theta)$}
\label{AppendQ}
Let ${\bf Z}_i$ denote the latent cluster assignments $Z_{i1},\ldots,Z_{iT_i}$.  Then we have
{\small
\begin{align}\nonumber
&Q(\Theta,\widehat\Theta) & \\ \nonumber
&=\sum_{{\bf Z}_1,\ldots,{\bf Z}_n}\sum_{i=1}^n\left[ \log(\alpha_{Z_{it}}\pi({\bf X}_{i1}|Z_{i1},\Theta))+\sum_{t=2}^{T_i}\log(\beta_{Z_{i(t-1)}Z_{it}}\pi({\bf X}_{it}|{\bf X}_{i(t-1)},Z_{it},\Theta))\right] \prod_{j=1}^n  \mathbb{P}({\bf Z}_j|{\cal X}_j,\widehat\Theta) & \\ \nonumber
&=\sum_{{\bf Z}_1,\ldots,{\bf Z}_n} \sum_{i=1}^n\sum_{\ell_1=1}^K\cdots\sum_{\ell_{T_i}=1}^K1_{[Z_{i1}=\ell_1]}\cdots1_{[Z_{iT_i}=\ell_T]} & \\ \nonumber
&\cdot \left[ \log(\alpha_{\ell_1}\pi({\bf X}_{i1}|Z_{i1}=\ell_1,\Theta))+\sum_{t=2}^{T_i}\log(\beta_{\ell_{t-1}\ell_t}\pi({\bf X}_{it}|{\bf X}_{i(t-1)},Z_{it}=\ell_t,\Theta))\right] \prod_{j=1}^n \mathbb{P}({\bf Z}_j|{\cal X}_j,\widehat\Theta) & \\ \nonumber
&= \sum_{i=1}^n\sum_{\ell_1=1}^K\cdots\sum_{\ell_{T_i}=1}^K  \left[ \log(\alpha_{\ell_1}\pi({\bf X}_{i1}|Z_{i1}=\ell_1,\Theta)) + \sum_{t=2}^{T_i}\log(\beta_{\ell_{t-1}\ell_t}\pi({\bf X}_{it}|{\bf X}_{i(t-1)},Z_{it}=\ell_t,\Theta))\right] & \\
&  \hspace{1pc}\cdot\sum_{{\bf Z}_1,\ldots,{\bf Z}_n} 1_{[Z_{i1}=\ell_1]}\cdots1_{[Z_{iT_i}=\ell_{T_i}]}
\prod_{j=1}^n \mathbb{P}({\bf Z}_j|{\cal X}_j,\widehat\Theta) .&
\end{align}
}
We can simplify this last expression by noting that for a fixed $i$,
\begin{eqnarray}\nonumber
&& \sum_{{\bf Z}_1,\ldots,{\bf Z}_n} 1_{[Z_{i1}=\ell_1]}\cdots1_{[Z_{iT_i}=\ell_T]}
\prod_{j=1}^n \mathbb{P}({\bf Z}_j|{\cal X}_j,\widehat\Theta) \\ \nonumber
&=&\left[\prod_{j\neq i} \sum_{{\bf Z}_j}\mathbb{P}({\bf Z}_j|{\cal X}_j,\widehat\Theta)\right]\mathbb{P}(Z_{i1}=\ell_1,\ldots,Z_{iT_i}=\ell_{T_i}|X_i,\widehat\Theta) \\
&=& \mathbb{P}(Z_{i1}=\ell_1|{\cal X}_i,\widehat\Theta)\prod_{t=1}^{T_i}\mathbb{P}(Z_{it}=\ell_t|Z_{i(t-1)}=\ell_{t-1},{\cal X}_i,\widehat\Theta).
\end{eqnarray}
Thus we can rewrite $Q$ in the more tractable form
\begin{align}\nonumber
&Q(\Theta,\widehat\Theta) & \\ \nonumber
&= \sum_{i=1}^n \sum_{\ell_1=1}^K\cdots\sum_{\ell_{T_i}=1}^K \left[ \log(\alpha_{\ell_1})+\sum_{t=2}^{T_i}\log(\beta_{\ell_{t-1}\ell_t})\right]\mathbb{P}(Z_{i1}=\ell_1|{\cal X}_i,\widehat\Theta)\prod_{s=2}^{T_i}\mathbb{P}(Z_{is}=\ell_s|Z_{i(s-1)}=\ell_{s-1},{\cal X}_i,\widehat\Theta) & \\ \nonumber
&+ \sum_{i=1}^n \sum_{\ell_1=1}^K\cdots\sum_{\ell_{T_i}=1}^K \left[ \log(\pi({\bf X}_{i1}|Z_{i1}=\ell_1,\Theta))+\sum_{t=2}^{T_i}\log(\pi({\bf X}_{it}|{\bf X}_{i(t-1)},Z_{it}=\ell_t,\Theta))\right] &\\
&\hspace{9pc}\cdot \mathbb{P}(Z_{i1}=\ell_1|{\cal X}_i,\widehat\Theta)\prod_{s=2}^{T_i}\mathbb{P}(Z_{is}=\ell_s|Z_{i(s-1)}=\ell_{s-1},{\cal X}_i,\widehat\Theta). &
\end{align}

\vspace{-1pc}
\section{Computing Marginal and Conditional Distributions of $Z_{it}$}
\label{AppendZdists}
As in the main text, the subscript $i$ is suppressed.  Note that the dependence on $\Theta$ has also been suppressed for ease of notation.
We start by noticing that
\begin{eqnarray}
\mathbb{P}(Z_{T}|Z_{T-1},{\cal X})\propto\pi({\cal X}|Z_{T},Z_{T-1})\mathbb{P}(Z_{T}|Z_{T-1})
\propto\beta_{Z_{T-1}Z_{T}}\pi({\bf X}_{T}|{\bf X}_{T-1},Z_{T}).
\end{eqnarray}
We can then find the iterative relationship, for $2\leq t < T$,
\begin{align}\nonumber
&\mathbb{P}(Z_{t}|Z_{t-1},{\cal X}) & \\ \nonumber
&\propto\pi({\bf X}_{t},\ldots,{\bf X}_{T}|{\bf X}_{1},\ldots,{\bf X}_{t-1},Z_{t},Z_{t-1})\mathbb{P}(Z_{t}|Z_{t-1})&\\ \nonumber
&\propto\beta_{Z_{t-1}Z_{t}}\sum_{\ell_{t+1}}\cdots\sum_{\ell_{T}}\pi({\bf X}_{t},\ldots,{\bf X}_{T},Z_{t+1}=\ell_{t+1},\ldots,Z_{T}=\ell_{T}|{\bf X}_{t-1},Z_{t}) &\\ \nonumber
&\propto\beta_{Z_{t-1}Z_{t}}\pi({\bf X}_{t}|{\bf X}_{t-1},Z_{t}) & \\ \nonumber
&\cdot
\sum_{\ell_{t+1}}\Big(\beta_{Z_{t}\ell_{t+1}}\pi({\bf X}_{t+1}|{\bf X}_{t},Z_{t+1}=\ell_{t+1})\cdots
\sum_{\ell_{T}}\Big(\beta_{\ell_{T-1}\ell_{T}}\pi({\bf X}_{T}|{\bf X}_{T-1},Z_{T}=\ell_{T})\Big)\cdots\Big) &\\
&\propto\beta_{Z_{t-1}Z_{t}}\pi({\bf X}_{t}|{\bf X}_{t-1},Z_{t})
\sum_{\ell_{t+1}}q(Z_{t+1}=\ell_{t+1}|Z_{t}).&
\end{align}

\section{Deriving the Parameter Updates}
\label{AppendParms}
\subsection{Update $\boldsymbol\alpha$}
Letting $\lambda_{\alpha}$ be the Lagrange multiplier corresponding to the constraint $\sum_{\ell=1}^K\alpha_{\ell}=1$, we have
\begin{align}\nonumber
\frac{\partial Q}{\partial \alpha_h}&=\frac{\partial }{\partial \alpha_h}\left\{\sum_{i=1}^n \sum_{Z_{i1}=1}^K\cdots\sum_{Z_{iT_i}=1}^K \log(\alpha_{Z_{i1}})\mathbb{P}(Z_{i1}|{\cal X}_i,\widehat\Theta)\prod_{s=2}^{T_i}\mathbb{P}(Z_{is}|Z_{i(s-1)},{\cal X}_i,\widehat\Theta) -\lambda_{\alpha}\left( \sum_{\ell=1}^K\alpha_{\ell}-1 \right)  \right\}& \\
&=\frac{1}{\alpha_h}\sum_{i=1}^n\mathbb{P}(Z_{i1}=h|{\cal X}_i,\widehat\Theta)\prod_{s=2}^{T_i}\sum_{Z_{is}=1}^K\mathbb{P}(Z_{is}|Z_{i(s-1)},{\cal X}_i,\widehat\Theta)-\lambda_{\alpha},&
\end{align}
hence
\begin{equation}
\hat{\alpha}_h=\frac{1}{n}\sum_{i=1}^n\mathbb{P}(Z_{i1}=h|{\cal X}_i,\widehat\Theta).
\end{equation}
\subsection{Update $\beta_{hk}$}
Here we have $K$ Lagrange multipliers corresponding to the constraint on each row of the transition matrix, specifically that $\sum_{k=1}^K \beta_{jk}=1$ for $j=1,\ldots,K$.  Denoting these as $\lambda_j$, we have
\begin{eqnarray}\nonumber
\frac{\partial Q}{\partial \beta_{hk}}&=&\frac{\partial }{\partial \beta_{hk}}\left\{\sum_{i=1}^n \sum_{Z_{i1}=1}^K\cdots\sum_{Z_{iT_i}=1}^K\sum_{t=2}^{T_i} \log(\beta_{Z_{i(t-1)}Z_{it}})\mathbb{P}(Z_{i1}|{\cal X}_i,\widehat\Theta)\prod_{s=2}^{T_i}\mathbb{P}(Z_{is}|Z_{i(s-1)},{\cal X}_i,\widehat\Theta) \right. \\
&&\hspace{1pc} \left.-\sum_{j=1}^K\lambda_j\left( \sum_{m=1}^K\beta_{jm}-1 \right)  \right\}.
\end{eqnarray}
We can make this expression more tractable by noticing that, for any $i$ and $t\geq 2$, \footnotesize
\begin{align}\nonumber
&\hspace{-0.5pc}\sum_{Z_{i1}=1}^K\cdots\sum_{Z_{iT_i}=1}^K\log(\beta_{Z_{i(t-1)}Z_{it}})
\mathbb{P}(Z_{i1}|{\cal X}_i,\widehat\Theta)\prod_{s=2}^{T_i}\mathbb{P}(Z_{is}|Z_{i(s-1)},{\cal X}_i,\widehat\Theta)&\\ \nonumber
&\hspace{-1.5pc} =\sum_{Z_{i1}=1}^K\hspace{-0.25pc}\left(\hspace{-0.25pc}\mathbb{P}(Z_{i1}|{\cal X}_i,\widehat\Theta)\sum_{Z_{i2}=1}^K\hspace{-0.25pc}\left(\hspace{-0.25pc} \mathbb{P}(Z_{i2}|Z_{i1},{\cal X}_i,\widehat\Theta)\cdots\sum_{Z_{it}=1}^K\hspace{-0.25pc}\left(\hspace{-0.25pc} \log(\beta_{Z_{i(t-1)}Z_{it}})\mathbb{P}(Z_{it}|Z_{i(t-1)},{\cal X}_i,\widehat\Theta)\cdots\sum_{Z_{iT_i}=1}^K \mathbb{P}(Z_{iT_i}|Z_{i(T_i-1)},{\cal X}_i,\widehat\Theta) \right)\cdots\right)\right.&\\
&\hspace{-1.5pc} =\sum_{Z_{i(t-1)}=1}^K\mathbb{P}(Z_{i(t-1)}|{\cal X}_i,\widehat\Theta)\sum_{Z_{it}=1}^K \log(\beta_{Z_{i(t-1)}Z_{it}})\mathbb{P}(Z_{it}|Z_{i(t-1)},{\cal X}_i,\widehat\Theta).&
\end{align}\normalsize
Thus we can find that
\begin{eqnarray}
\frac{\partial Q}{\partial \beta_{hk}}&=&\frac{1}{\beta_{hk}}\sum_{i=1}^n\sum_{t=2}^{T_i}\mathbb{P}(Z_{i(t-1)}=h|{\cal X}_i,\widehat\Theta)\mathbb{P}(Z_{it}=k|Z_{i(t-1)}=h,{\cal X}_i,\widehat\Theta) - \lambda_h
\end{eqnarray}
and hence
\begin{equation}
\hat{\beta}_{hk}=\frac{\sum_{i=1}^n\sum_{t=2}^{T_i}\mathbb{P}(Z_{i(t-1)}=h|{\cal X}_i,\widehat\Theta)\mathbb{P}(Z_{it}=k|Z_{i(t-1)}=h,{\cal X}_i,\widehat\Theta)}{\sum_{i=1}^n\sum_{t=2}^{T_i}\mathbb{P}(Z_{i(t-1)}=h|{\cal X}_i,\widehat\Theta)}.
\end{equation}

\subsection{Update $\lambda$, $\boldsymbol\mu_k$ and $\Sigma_k$}
Before we proceed, we need a few preliminaries.
The partial derivatives of (\ref{transition1}) and (\ref{transition2}) with respect to $\boldsymbol\mu_k$, $k=1,\ldots,K$, are
\begin{eqnarray}
\frac{\partial}{\partial \boldsymbol\mu_k} \log(\pi({\bf X}_{i1}|k,\Theta)) =\frac{\partial}{\partial \boldsymbol\mu_k}\left(-\frac{1}{2}\right)\left[ ({\bf X}_{i1}-\boldsymbol\mu_k)'\Sigma_k^{-1}({\bf X}_{i1}-\boldsymbol\mu_k) \right]
=\Sigma_k^{-1}{\bf X}_{i1}-\Sigma_k^{-1}\boldsymbol\mu_k
\end{eqnarray}
\begin{eqnarray}\nonumber
&&\frac{\partial}{\partial \boldsymbol\mu_k} \log(\pi({\bf X}_{it}|{\bf X}_{i(t-1)},k,\Theta)) \\ \nonumber
&=&\frac{\partial}{\partial \boldsymbol\mu_k}(-\frac{1}{2})\left[ ({\bf X}_{it}-(1-\lambda){\bf X}_{i(t-1)}-\lambda\boldsymbol\mu_k)'\Sigma_k^{-1}({\bf X}_{it}-(1-\lambda){\bf X}_{i(t-1)}-\lambda\boldsymbol\mu_k) \right] \\
&=&\lambda\Sigma_k^{-1}({\bf X}_{it}-(1-\lambda){\bf X}_{i(t-1)})-\lambda^2\Sigma_k^{-1}\boldsymbol\mu_k,
\end{eqnarray}
with respect to $\Sigma_k^{-1}$, $k=1,\ldots,K$, are
\begin{eqnarray}\nonumber
\frac{\partial}{\partial \Sigma_k^{-1}} \log(\pi({\bf X}_{i1}|k,\Theta))
&=&\frac{\partial}{\partial \Sigma_k^{-1}}\left[ \frac{1}{2}\log|\Sigma_k^{-1}|-\frac{1}{2}tr(\Sigma_k^{-1}({\bf X}_{it}-\boldsymbol\mu_k)({\bf X}_{it}-\boldsymbol\mu_k)') \right] \\ \nonumber
&=&\Sigma_k-\frac{1}{2}diag(\Sigma_k)-({\bf X}_{it}-\boldsymbol\mu_k)({\bf X}_{it}-\boldsymbol\mu_k)'+\frac{1}{2}diag(({\bf X}_{it}-\boldsymbol\mu_k)({\bf X}_{it}-\boldsymbol\mu_k)') \\
&=&\Sigma_k\circ \Big(\frac{1}{2}I_p\Big)-({\bf X}_{it}-\boldsymbol\mu_k)({\bf X}_{it}-\boldsymbol\mu_k)'\circ \Big(\frac{1}{2}I_p\Big),
\end{eqnarray}
where $\circ$ is the Hadamard product, and similarly
\begin{eqnarray}\nonumber
&&\frac{\partial}{\partial \Sigma_k^{-1}} \log(\pi({\bf X}_{it}|{\bf X}_{i(t-1)},k,\Theta))\\
&=& \Sigma_k\circ \Big(\frac{1}{2}I_p\Big)-\big({\bf X}_{it}-\lambda\boldsymbol\mu_k-(1-\lambda){\bf X}_{i(t-1)}\big)\big({\bf X}_{it}-\lambda\boldsymbol\mu_k-(1-\lambda){\bf X}_{i(t-1)}\big)'\circ \Big(\frac{1}{2}I_p\Big),
\end{eqnarray}
and with respect to $\lambda$ is
\begin{eqnarray}\nonumber
&&\frac{\partial}{\partial \lambda} \log(\pi({\bf X}_{it}|{\bf X}_{i(t-1)},k,\Theta)) \\ \nonumber
&=&\frac{\partial}{\partial \lambda}(-\frac{1}{2})\left[ ({\bf X}_{it}-\lambda(\boldsymbol\mu_k-{\bf X}_{i(t-1)})-{\bf X}_{i(t-1)})'\Sigma_k^{-1}({\bf X}_{it}-\lambda(\boldsymbol\mu_k-{\bf X}_{i(t-1)})-{\bf X}_{i(t-1)}) \right] \\
&=&({\bf X}_{it}-{\bf X}_{i(t-1)})'\Sigma_k^{-1}(\boldsymbol\mu_k-{\bf X}_{i(t-1)})-\lambda(\boldsymbol\mu_k-{\bf X}_{i(t-1)})'\Sigma_k^{-1}(\boldsymbol\mu_k-{\bf X}_{i(t-1)}).
\end{eqnarray}
To make the form of $Q(\Theta,\widehat\Theta)$ more tractable, we notice that for any $i$,
\begin{eqnarray}\nonumber
&&\sum_{Z_{i1}=1}^K\cdots\sum_{Z_{iT_i}=1}^K\log(\pi({\bf X}_{i1}|Z_{i1},\Theta)) \mathbb{P}(Z_{i1}|{\cal X}_j,\widehat\Theta)\prod_{s=2}^{T_i}\mathbb{P}(Z_{is}|Z_{i(s-1)},{\cal X}_i,\widehat\Theta) \\ \nonumber
&=&\sum_{Z_{i1}=1}^K\log(\pi({\bf X}_{i1}|Z_{i1},\Theta)) \mathbb{P}(Z_{i1}|{\cal X}_j,\widehat\Theta)
\prod_{s=2}^{T_i}\sum_{Z_{is}=1}^K \mathbb{P}(Z_{is}|Z_{i(s-1)},{\cal X}_i,\widehat\Theta) \\
&=&\sum_{Z_{i1}=1}^K\log(\pi({\bf X}_{i1}|Z_{i1},\Theta)) \mathbb{P}(Z_{i1}|{\cal X}_j,\widehat\Theta) ,
\end{eqnarray}
and for any $i$ and $t\geq 2$,
\begin{align}\nonumber
&\sum_{Z_{i1}=1}^K\cdots\sum_{Z_{iT_i}=1}^K\log(\pi({\bf X}_{it}|{\bf X}_{i(t-1)},Z_{it},\Theta)) \mathbb{P}(Z_{i1}|{\cal X}_j,\widehat\Theta)\prod_{s=2}^{T_i}\mathbb{P}(Z_{is}|Z_{i(s-1)},{\cal X}_i,\widehat\Theta) &\\ \nonumber
=&\sum_{Z_{i1}=1}^K\cdots\sum_{Z_{it}=1}^K \mathbb{P}(Z_{i1}|{\cal X}_j,\widehat\Theta)\left[\prod_{s=2}^{t-1}\mathbb{P}(Z_{is}|Z_{i(s-1)},{\cal X}_i,\widehat\Theta)\right]
\log(\pi({\bf X}_{it}|{\bf X}_{i(t-1)},Z_{it},\Theta))\mathbb{P}(Z_{it}|Z_{i(t-1)},{\cal X}_i,\widehat\Theta) &\\ \nonumber
&\hspace{1pc}\cdot \left[ \prod_{u=t+1}^{T_i}\sum_{Z_{iu}=1}^K \mathbb{P}(Z_{iu}|Z_{i(u-1)},{\cal X}_i,\widehat\Theta) \right] &\\
=&\sum_{Z_{it}=1}^K\log(\pi({\bf X}_{it}|{\bf X}_{i(t-1)},Z_{it},\Theta))\mathbb{P}(Z_{it}|{\cal X}_i,\widehat\Theta).&
\end{align}

With the above it is not difficult to find, for each distribution parameter, the value which maximizes $Q(\Theta,\widehat\Theta)$ where the other parameters are fixed, hence finding the solutions to the coordinate ascent approach.  The solutions are found as follows.  The derivative of $Q(\Theta,\widehat\Theta)$ with respect to $\lambda$ is
\begin{eqnarray}\nonumber
\frac{\partial Q}{\partial \lambda}&=&\sum_{i=1}^n\sum_{t=2}^{T_i}\sum_{Z_{it}=1}^K\mathbb{P}(Z_{it}|{\cal X}_i,\widehat\Theta)\\
&& \cdot \left[({\bf X}_{it}-{\bf X}_{i(t-1)})'\Sigma_k^{-1}(\boldsymbol\mu_k-{\bf X}_{i(t-1)})-\lambda(\boldsymbol\mu_k-{\bf X}_{i(t-1)})'\Sigma_k^{-1}(\boldsymbol\mu_k-{\bf X}_{i(t-1)})\right].
\end{eqnarray}
Hence the update for $\lambda$ is
\begin{equation}
\hat{\lambda}=\frac{\sum_{i=1}^n\sum_{t=2}^{T_i}\sum_{Z_{it}=1}^K\mathbb{P}(Z_{it}|{\cal X}_i,\widehat\Theta)({\bf X}_{it}-{\bf X}_{i(t-1)})'\Sigma_k^{-1}(\boldsymbol\mu_k-{\bf X}_{i(t-1)})}
{\sum_{j=1}^n\sum_{s=2}^{T_j}\sum_{Z_{js}=1}^K\mathbb{P}(Z_{js}|{\cal X}_j,\widehat\Theta)(\boldsymbol\mu_k-{\bf X}_{j(s-1)})'\Sigma_k^{-1}(\boldsymbol\mu_k-{\bf X}_{j(s-1)})}.
\end{equation}
The derivative of $Q(\Theta,\widehat\Theta)$ with respect to $\boldsymbol\mu_k$, $k=1,\ldots,K$, is
\begin{eqnarray}\nonumber
\frac{\partial Q}{\partial \boldsymbol\mu_k}&=&\sum_{i=1}^n \biggl\{  \mathbb{P}(Z_{i1}=k|{\cal X}_i,\widehat\Theta)\left[\Sigma_k^{-1}{\bf X}_{i1}-\Sigma_k^{-1}\boldsymbol\mu_k  \right]  \\
&&+\sum_{t=2}^{T_i}\mathbb{P}(Z_{it}=k|{\cal X}_i,\widehat\Theta)\left[ \lambda\Sigma_k^{-1}({\bf X}_{it}-(1-\lambda){\bf X}_{i(t-1)})-\lambda^2\Sigma_k^{-1}\boldsymbol\mu_k  \right] \biggr\}.
\end{eqnarray}
Hence the update for $\boldsymbol\mu_k$ is
\begin{equation}
\hat{\boldsymbol\mu}_k=\frac{\sum_{i=1}^n \biggl\{  \mathbb{P}(Z_{i1}=k|{\cal X}_i,\widehat\Theta){\bf X}_{i1}+\lambda \sum_{t=2}^{T_i}\mathbb{P}(Z_{it}=k|{\cal X}_i,\widehat\Theta)({\bf X}_{it}-(1-\lambda){\bf X}_{i(t-1)}) \biggr\}}{\sum_{i=1}^n \biggl\{  \mathbb{P}(Z_{i1}=k|{\cal X}_i,\widehat\Theta) +\lambda^2\sum_{t=2}^{T_i}\mathbb{P}(Z_{it}=k|{\cal X}_i,\widehat\Theta) \biggr\}}.
\end{equation}
Let ${\bf A}_{ik}={\bf X}_{i1}-\boldsymbol\mu_k$ and ${\bf B}_{itk}={\bf X}_{it}-\lambda\boldsymbol\mu_k-(1-\lambda){\bf X}_{i(t-1)}$.  Then the derivative of $Q(\Theta,\widehat\Theta)$ with respect to $\Sigma_k^{-1}$, $k=1,\ldots,K$, is
\begin{align} \nonumber
\frac{\partial Q}{\partial \Sigma_k^{-1}}= &\sum_{i=1}^n \biggl\{  \mathbb{P}(Z_{i1}=k|{\cal X}_i,\widehat\Theta)\left[\Sigma_k\circ\Big(\frac{1}{2}I_p\Big)-{\bf A}_{ik}{\bf A}_{ik}'\circ\Big(\frac{1}{2}I_p\Big)  \right]  &\\
&+\sum_{t=2}^{T_i}\mathbb{P}(Z_{it}=k|{\cal X}_i,\widehat\Theta)\left[ \Sigma_k\circ\Big(\frac{1}{2}I_p\Big)-{\bf B}_{itk}{\bf B}_{itk}'\circ\Big(\frac{1}{2}I_p\Big)  \right] \biggr\}.&
\end{align}
Hence the update for $\Sigma_k$ can be found as
\begin{align}\nonumber
\widehat{\Sigma}_k&= \frac{\sum_{i=1}^n \biggl\{  \mathbb{P}(Z_{i1}=k|{\cal X}_i,\widehat\Theta){\bf A}_{ik}{\bf A}_{ik}'+\sum_{t=2}^{T_i}\mathbb{P}(Z_{it}=k|{\cal X}_i,\widehat\Theta){\bf B}_{itk}{\bf B}_{itk}'\biggr\}}{\sum_{i=1}^n \biggl\{  \mathbb{P}(Z_{i1}=k|{\cal X}_i,\widehat\Theta) +\sum_{t=2}^{T_i}\mathbb{P}(Z_{it}=k|{\cal X}_i,\widehat\Theta) \biggr\}}.&
\end{align}

\section{Details on Variables Measured on MCs}
\label{AppendMC}
Our variables include the number of district offices operated by the MC, the proportion of legislative staff assigned to the district (rather than Capitol Hill), the number of bill introductions, cosponsorships, and amendments made by the MC, the number of one-minute speeches the MC gives on the House floor, the number of editorials and opinion pieces he or she writes for state and national papers, the total amount of campaign money he or she raises, the total amount of money he or she contributes to the party campaign committee (the Democratic Congressional Campaign Committee, which collects and redistributes funds for election campaigns), the total amount of money he or she contributes directly to colleagues, the percentage of the time he or she votes with the party on party votes (defined as votes where a majority of one party votes against a majority of the other), the percentage of the time he or she votes with the  party leadership on leadership votes (defined as votes where the leadership of one party votes one way and the leadership of the other party the other), the number of issue areas (out of eighteen) in which he or she introduces legislative bills, the proportion of his or her cosponsorships that are bipartisan (i.e., for measures introduced by an MC from the Republican party), and the percentage of his or her introduced bills that are referred to a committee on which he or she sits.

Each index used in the clustering algorithm was constructed in the following way. First, each of the raw variables described above was standardized at each time point to eliminate global temporal patterns or shifts. Second, each variable was associated with one index (existing literature on congressional behavior pointed us to which variables should fall under each index). Finally, each index was constructed by averaging the associated variables. As an example, suppose that there are three raw variables, $X_{it}^{(1)}$, $X_{it}^{(2)}$, and $X_{it}^{(3)}$, associated with index one. Then the first index $X_{it1}$ (the first entry in ${\bf X}_{it}$) is computed as
\begin{equation*}
X_{it1}=\frac13\left[
\frac{X_{it}^{(1)}-\bar{X}^{(1)}_t}{s^{(1)}_t}+
\frac{X_{it}^{(2)}-\bar{X}^{(2)}_t}{s^{(2)}_t}+
\frac{X_{it}^{(3)}-\bar{X}^{(3)}_t}{s^{(3)}_t}
\right],
\end{equation*}
for each $i$ and $t$, where $\bar{X}^{(\ell)}_t$ and $s^{(\ell)}_t$ are the sample mean and sample standard deviation respectively of $\{X_{it}^{(\ell)}\}_{i=1}^n$, $\ell=1,2,3$.



\begin{thebibliography}{3}
\newcommand{\enquote}[1]{``#1''}
\expandafter\ifx\csname natexlab\endcsname\relax\def\natexlab#1{#1}\fi

\bibitem[{Anderlucci and Viroli(2014)}]{anderlucci2014covariance}
Anderlucci, L. and Viroli, C. (2014), \enquote{Covariance Pattern Mixture
  Models for Multivariate Longitudinal Data with Application to the Health and
  Retirement Study,} \textit{arXiv preprint arXiv:1401.1301}.

\bibitem[{De~la Cruz-Mes{\'\i}a et~al.(2008)De~la Cruz-Mes{\'\i}a, Quintana,
  and Marshall}]{de2008model}
De~la Cruz-Mes{\'\i}a, R., Quintana, F.~A., and Marshall, G. (2008),
  \enquote{Model-Based Clustering for Longitudinal Data,} \textit{Computational
  Statistics \& Data Analysis}, 52, 1441--1457.

\bibitem[{Fr{\"u}hwirth-Schnatter(2006)}]{fruhwirth2006finite}
Fr{\"u}hwirth-Schnatter, S. (2006), \textit{Finite Mixture and Markov Switching
  Models: Modeling and Applications to Random Processes}, Springer.

\end{thebibliography}


\begin{thebibliography}{11}
\bibitem{} Anderlucci, L. and Viroli, C. (2014). Covariance pattern mixture models for multivariate longitudinal data with application to the health and retirement study. {\it arXiv preprint arXiv:1401.1301.}
\bibitem[]{2} Banfield, J. D. and Raftery, A.E. (1993). Model-Based Gaussian and Non-Gaussian Clustering. {\it Biometrics} {\bf 49}, 803-821.
\bibitem[]{3}Chung, H., Park, Y., and Lanza, S. T. (2005). Latent transition analysis with covariates: Pubertal timing and substance use behaviours in adolescent females. {\it Statistics in Medicine} {\bf 24}, 2895-2910.
\bibitem[]{4} Collins, L. M., Graham, J. W., Rousculp, S. S., Fidler, P.L., Pan, J., and Hansen, W.B. (1994). Latent transition analysis and how it can address prevention research questions. {\it NIDA Research Monograph} {\bf 142}, 81-111.
\bibitem[]{5} Collins, L. M., and Wugalter, S. E. (1992). Latent class models for stage-sequential dynamic latent variables. {\it Multivariate Behavioral Research} {\bf 27}, 131-157.
\bibitem[]{6} De la Cruz-Mes{\'\i}a, R. and Quintana, F. A. and Marshall, G. (2008). Model-based clustering for longitudinal data. {\it Computational Statistics \& Data Analysis} {\bf 52}, 1441-1457.
\bibitem[]{7}Dempster, A. P., Laird, N. M., and Rubin, D. B. (1977). Maximum likelihood from incomplete data via the EM algorithm. {\it Journal of the Royal Statistical Society B} {\bf 39}, 1-38.
\bibitem[]{8} Fr{\"u}hwirth-Schnatter, S. (2006). {\it Finite Mixture and Markov Switching Models: Modeling and Applications to Random Processes}. Springer.
\bibitem[]{9} Gaffney, S. and Smyth, P. (1999). Trajectory clustering with mixtures of regression models. In {\it Proceedings of the Fifth ACM SIGKDD International Conference on Knowledge Discovery and Data Mining}, ACM, 63-72.
\bibitem[]{10} Genolini, C. and Falissard, B. (2010). KmL: k-means for longitudinal data. {\it Computational Statistics} {\bf 25}, 317-328.
\bibitem[]{11}Graham, J. W., Collins, L. M., Wugalter, S. E., Chung, N. K. and Hansen, W. B. (1991). Modeling transitions in latent stage-sequential processes: a substance use prevention example. {\it Journal of Consulting and Clinical Psychology} {\bf 59}, 48-57.
\bibitem[]{12}Hubert, L. and Arabie, P. (1985). Comparing partitions. {\it Journal of Classification} {\bf 2}, 193-218.
\bibitem[]{13}Lou, S., Jiang, J. and Keng, K. (1993). Clustering objects generated by linear regression models. {\it Journal of the American Statistical Association}. {\bf 88}, 1356-1362.
\bibitem[]{14}Luan, Y. and Li, H. (2003). Clustering of time-course gene expression data using a mixed-effects model with B-splines. {\it Bioinformatics} {\bf 19}, 474-482.
\bibitem[]{15}McNicholas, P. D. and Murphy, T. B. (2010). Model-based clustering of longitudinal data. {\it Canadian Journal of Statistics} {\bf 38}, 153-168.
\bibitem[]{16}Meil{\u{a}}, M. (2003). Comparing clusterings by the variation of information. In {\it Learning Theory and Kernel Machines}, 173-189. Springer.
\bibitem[]{17}Ray, S. and Mallick, B. (2006). Functional clustering by Bayesian wavelet methods. {\it Journal of the Royal Statistical Society B} {\bf 68}, 305-332.
\bibitem[]{18}Rousseeuw, P. J. (1987). Silhouettes: a graphical aid to the interpretation and validation of cluster analysis. {\it Journal of Computational and Applied Mathematics} {\bf 20}, 53-65.
\bibitem[]{19}Scott, S. L. James, G. M., and Sugar, C. A. (2005). Hidden Markov models for longitudinal comparisons. {\it Journal of the American Statistical Association} {\bf 24}, 179-207.
\bibitem[]{20}Vermunt, J. K., Langeheine, R., and Bockenholt, U. (1999). Discrete-time discrete-state latent Markov models with time-constant and time-varying covariates. {\it Journal of Educational and Behavioral Statistics} {\bf 24}, 179-207.
\bibitem[]{21}Wu, C. (1983). On the convergence properties of the EM algorithm. {\it The Annals of Statistics} {\bf 11}, 95-103.
\end{thebibliography}

\end{document}